\documentclass[onecolumn,amsmath,showpacs,nofootinbib,12pt]{revtex4-2}
\usepackage{graphicx}
\usepackage{dcolumn}
\usepackage{bm}
\usepackage{color} 
\usepackage{slashed}
\begin{document}
\newcommand{\hs}{\hspace*{0.5cm}}
\newcommand{\vs}{\vspace*{0.5cm}}
\newcommand{\be}{\begin{equation}}
\newcommand{\ee}{\end{equation}}
\newcommand{\bea}{\begin{eqnarray}}
\newcommand{\eea}{\end{eqnarray}}
\newcommand{\ben}{\begin{enumerate}}
\newcommand{\een}{\end{enumerate}}
\newcommand{\bde}{\begin{widetext}}
\newcommand{\ede}{\end{widetext}}
\newcommand{\nn}{\nonumber}
\newcommand{\crn}{\nonumber \\}
\newcommand{\Tr}{\mathrm{Tr}}
\newcommand{\non}{\nonumber}
\newcommand{\noi}{\noindent}
\newcommand{\al}{\alpha}
\newcommand{\la}{\lambda}
\newcommand{\bet}{\beta}
\newcommand{\ga}{\gamma}
\newcommand{\va}{\varphi}
\newcommand{\om}{\omega}
\newcommand{\pa}{\partial}
\newcommand{\+}{\dagger}
\newcommand{\fr}{\frac}
\newcommand{\bc}{\begin{center}}
\newcommand{\ec}{\end{center}}
\newcommand{\Ga}{\Gamma}
\newcommand{\de}{\delta}
\newcommand{\De}{\Delta}
\newcommand{\ep}{\epsilon}
\newcommand{\varep}{\varepsilon}
\newcommand{\ka}{\kappa}
\newcommand{\La}{\Lambda}
\newcommand{\si}{\sigma}
\newcommand{\Si}{\Sigma}
\newcommand{\ta}{\tau}
\newcommand{\up}{\upsilon}
\newcommand{\Up}{\Upsilon}
\newcommand{\ze}{\zeta}
\newcommand{\ps}{\psi}
\newcommand{\Ps}{\Psi}
\newcommand{\ph}{\phi}
\newcommand{\vph}{\varphi}
\newcommand{\Ph}{\Phi}
\newcommand{\Om}{\Omega}
\newcommand{\AdrHEPC}{Phenikaa Institute for Advanced Study and Faculty of Basic Science, Phenikaa University, Yen Nghia, Ha Dong, Hanoi 100000, Vietnam}

\title{Scotoseesaw model implied by dark right-handed neutrinos} 

\author{Phung Van Dong} 
\email{dong.phungvan@phenikaa-uni.edu.vn (corresponding author)}
\author{Duong Van Loi} 
\email{loi.duongvan@phenikaa-uni.edu.vn}
\affiliation{\AdrHEPC} 

\date{\today}

\begin{abstract}
We find a dark gauge symmetry $U(1)_D$ that transforms nontrivially only for three right-handed neutrinos, $\nu_{1,2,3R}$. The anomaly cancellation demands that they have a dark charge $D=0,-1,+1$ assigned to $\nu_{1,2,3R}$, respectively. The dark charge is broken by two units down to a dark parity, i.e. $U(1)_D\to P_D=(-1)^D$, which stabilizes a dark matter candidate. Interestingly, the model manifestly supplies neutrino masses via joint seesaw and scotogenic mechanisms, called scotoseesaw.    
\end{abstract} 

\maketitle

\section{Motivation}

The standard model arranges left-handed fermions in doublets, $l_{aL}\equiv (\nu_{aL},e_{aL})$ and $q_{aL} \equiv (u_{aL},d_{aL})$, while it puts right-handed fermion partners in singlets, $e_{aR}$, $u_{aR}$, and $d_{aR}$, where $a=1,2,3$ is a family index. Since the standard model contains neither sterile right-handed neutrino singlets ($\nu_{aR}$) nor lepton number violating interactions, it implies massless neutrinos, in contrast with the experiment \cite{kajita,mcdonald}.

If the right-handed neutrinos $\nu_{aR}$ are included according to their respective families, they possess a Lagrangian $\mathcal{L}\supset x_{ab}\bar{l}_{aL} \tilde{H}\nu_{bR}-\fr 1 2 M_{ab} \bar{\nu}^c_{aR}\nu_{bR}+H.c.$, where $\tilde{H}=i\sigma_2 H^*$ relates to the standard model Higgs doublet $H=(H^+,H^0)$, $b=1,2,3$ denotes a family index as $a$ does, and the superscript $^c$ indicates charge conjugation. After electroweak symmetry breaking, Dirac neutrino masses that link $\nu_{L}$'s with $\nu_{R}$'s arise, such as $m_{ab}=-x_{ab} v/\sqrt{2}$, proportional to the weak scale $v=\sqrt{2}\langle H^0\rangle=246$ GeV, whereas the right-handed neutrinos possibly obtain large Majorana masses $M$'s as coupled to $\nu_R$'s by themselves, obeying $M\gg m$. This leads to a canonical seesaw mechanism producing appropriate small neutrino masses as suppressed by $m_\nu\simeq -m^2/M$ \cite{Minkowski,Yanagida,Gell-Mann,Mohapatra,valle}. However, the canonical seesaw mechanism in itself does not manifestly explain the existence of dark matter, similar to the standard model \cite{bertone,arcadi}. 

If the right-handed neutrinos $\nu_{R}$'s are odd under an exact symmetry $Z_2$ (whereas, all the standard model particles are even under this group), the Dirac masses $m$'s, or exactly the coupling $x_{ab}$, are suppressed, while the Majorana masses $M$'s are allowed. Additionally, the left-handed neutrinos $\nu_L$'s now couple to $\nu_R$'s via an extra inert Higgs doublet $\eta=(\eta^0,\eta^-)$ that is odd under $Z_2$ too, given through a Lagrangian $\mathcal{L}\supset y_{ab}\bar{l}_{aL}\eta \nu_{bR}-\fr 1 2 M_{ab}\bar{\nu}^c_{aR}\nu_{bR}+H.c.$, where the Majorana masses $M$'s must be included. In this way, appropriate neutrino masses are radiatively induced by the dark fields $\nu_R$'s and $\eta^0$'s that run in the loop, called the scotogenic mechanism \cite{tao,ma}. This approach is well motivated since the lightest of the dark fields $\nu_R$'s and $\eta^0$'s is stabilized by $Z_2$, responsible for dark matter. However, the existence of the $Z_2$ symmetry---which is {\it ad hoc} input---is a mystery. 

Because every discrete symmetry, such as the $Z_2$ group, would arise from a more fundamental gauge symmetry, as proven by Krauss and Wilczek \cite{krauss} long ago, it is worthwhile to look for the simplest gauge symmetry, being perhaps a continuous Abelian group, which transforms nontrivially for the right-handed neutrinos and the inert Higgs doublet, similar to its residual $Z_2$. This $Z_2$ symmetry may be a residual matter parity resulting from the baryon-minus-lepton number ($B-L$) gauge breaking, say $P_M=(-1)^{3(B-L)+2s}$, where $(-1)^{2s}$ is the spin-parity conserved, included for convenience. In this kind of theory, $\nu_R$'s and $\eta$ actually possess $B-L$ charges to be even and odd, respectively, opposite to the property of particles in the standard model. For instance, the coupling $y_{ab}\bar{l}_{aL}\eta\nu_{bR}$ is allowed for $B-L=0$ for $\nu_R$'s and $-1$ for $\eta$. In this case, the scotogenic scheme works because of the matter parity conservation, in which $P_M=-1$ for these new fields, while $P_M=1$ for the usual particles (see, e.g., \cite{ma1,valle1}). However, the anomalies relevant to $B-L$ are not free, since $[B-L](\nu_R)\neq [B-L](\nu_L)$. The anomaly cancellation requires extra chiral fermions beyond the conventional right-handed neutrinos, $\nu_{1,2,3R}$. 

Alternatively, the $Z_2$ group may result from breaking of a dark gauge charge, called $D$, for which all the standard model particles have no dark charge, $D=0$, while new (dark) fields actually take $D\neq 0$. The anomaly cancellation associated with $D$ demands that new fermions would be vectorlike under $D$, i.e. Dirac fermions. If the counterparts of $\nu_L$'s are three Dirac fermions $N_a$ charged under $D=-1$ and coupled to normal matter via inert Higgs doublets $\eta,\chi$ with $D=1,-1$, respectively, such as $\mathcal{L}\supset y_{ab}\bar{l}_{aL}\eta N_{bR}+z_{ab}\bar{l}_{aL}\chi N^c_{bL} - M_{ab}\bar{N}_{aL} N_{bR}+H.c.$, this leads to a scotogenic scheme for neutrino mass generation in which all new neutral fields $N_{L,R}$ and $\eta^0,\chi^0$ run in the loop \cite{dcz2}.\footnote{Let us call the reader's attention to a non-Abelian dark charge extension that also leads to scotogenic neutrino mass generation, e.g., \cite{dlh}.} Although the number of Dirac fermions $N$'s from the outset is three, we need at least two of these fields, i.e. four chiral fermions $N_L$'s and $N_R$'s, in order to successfully explain the neutrino oscillation data. Again, the number of chiral fermions acquired is beyond the conventional three $\nu_R$'s.     

According to the above review, to our best knowledge, most gauge extensions of the scotogenic setup that properly stabilize a dark matter candidate require extra degrees of fermions of freedom beyond the conventional three right-handed neutrinos---the counterparts of usual neutrinos. Although such proposals are interesting, this work will not examine them further. Instead, we look for a novel Abelian gauge symmetry that governs just three right-handed neutrinos, as in the original scotogenic setup, for neutrino mass generation and dark matter stability. Interestingly, the anomaly cancellation requires a right-handed neutrino ($\nu_{1R}$) with a new charge $D=0$, while two right-handed neutrinos ($\nu_{2,3R}$) obtain the new charges $D=-1,+1$, respectively, which properly form a Dirac fermion. The right-handed neutrino $\nu_{1R}$ couples to the usual lepton and Higgs doublets, producing a seesaw neutrino mass at the tree level, while the remaining $\nu_{2,3R}$ couple the usual lepton doublets to extra inert Higgs fields, inducing a scotogenic neutrino mass at the one-loop level. Given that these kinds of new physics are at the same scale, this proposal naturally solves a mass hierarchy between solar and atmospheric neutrinos by comparing tree- and loop-level contributions. Further, the lightest of $\nu_{2,3R}$ and neutral inert scalars is stabilized by a residual gauge symmetry, responsible for dark matter. That said, the new symmetry breaking implies neutrino mass generation and dark matter stability as interplay between a canonical seesaw and a scotogenic scheme, the so-called scotoseesaw \cite{kubo,valless}. 

Let us remind the reader that the simplest scotoseesaw set by two right-handed neutrinos was first studied in \cite{kubo} and later examined in \cite{valless} by {\it ad hoc} imposing an anomalous $U(1)$ group that is broken to a residual $Z_2$ symmetry or, alternatively, including a $Z_2$ symmetry from the outset, respectively. In this minimal scotoseesaw scheme, one right-handed neutrino is even under $Z_2$, responsible for tree-level seesaw neutrino mass generation, while another right-handed neutrino is odd under $Z_2$, responsible for radiative scotogenic neutrino mass generation, besides providing a potential dark matter candidate. Since the two right-handed neutrinos transform differently under $Z_2$, this gives rise to an interesting hypothesis that it may result from a $Z_8$ flavor symmetry \cite{valless1,extrassdt2} or a $A_4$ flavor symmetry \cite{extrass3,extrassdt1,extrass4}. The vacuum stability of $Z_2$ has been extensively studied in \cite{valless2}. Last, but not least, a surprising solution alternative to the above $B-L$ interpretation is that the anomalies relevant to this charge can be cancelled out for just three right-handed neutrinos $\nu_{1,2,3R}$ with $B-L=-4,-4,+5$, respectively \cite{blano}. In this way, both $\nu_{1,2R}$ are odd under the matter parity ($P_M=-1$) realizing a minimal scotogenic scheme \cite{dongsg}, while $\nu_{3R}$ sets a canonical seesaw, which results in a framework of the scotoseesaw with three right-handed neutrinos \cite{valless5}. Hence, our proposal (and that from $B-L$) indicates that a gauge completion of the scotoseesaw requires at least three right-handed neutrinos in order for it to work properly.      

The rest of this work is organized as follows. In Sec. \ref{pro}, we propose the model. In Sec.~\ref{pot}, we diagonalize the scalar potential. In Sec. \ref{neum}, we examine the neutrino mass. In Sec. \ref{darkm}, we investigate dark matter. In Sec. \ref{bounds}, we discuss theoretical and experimental constraints. We conclude this work and present an outlook in Sec.~\ref{concl}. Finally, Appendix \ref{appa} clarifies a radiative contribution to seesaw neutrino mass generation.   

\section{\label{pro} Proposal} 

This work for the first time assumes that the three right-handed neutrinos $\nu_{aR}(a=1,2,3)$ by themselves transform nontrivially under a dark gauge symmetry $U(1)_D$, i.e. each $\nu_{aR}$ is assigned to a dark charge $D_a$ according to $a=1,2,3$, whereas all the standard model fields possess $D=0$. The cancellation of $[\mathrm{Gravity}]^2 U(1)_D$ and $[U(1)_D]^3$ anomalies constrains $D_1+D_2+D_3=0$ and $D^3_1+D^3_2+D^3_3=0$, respectively. It leads to $D_1=0$, $D_2=-1$, and $D_3=+1$ as the unique solution, if indistinguishing $\nu_{1,2,3R}$ and normalizing the nonzero charge to unity, without loss of generality. Indeed, since $U(1)_D$ theory is invariant under a symmetry that transforms (i.e., rescaling) $D\to c D$ for charge and $g_D\to g_D/c$ for coupling, one can choose $c$ so that the charge normalization takes place. 

\begin{table}[h]
\bc
\begin{tabular}{lccccc}
\hline\hline 
Field & $SU(3)_C$ & $SU(2)_L$ & $U(1)_Y$ & $U(1)_{D}$ & $P_D$\\
\hline
$l_{aL} = \begin{pmatrix}
\nu_{aL}\\
e_{aL}\end{pmatrix}$ & 1 & 2 & $-1/2$ & $0$ & $+$\\
$\nu_{1R}$ & 1 & 1 & 0 & $0$ & $+$\\
$\nu_{2R}$ & 1 & 1 & 0 & $-1$ & $-$\\
$\nu_{3R}$ & 1 & 1 & 0 & $+1$ & $-$\\
$e_{aR}$ & 1 & 1 & $-1$ & $0$ & $+$\\
$q_{aL}= \begin{pmatrix}
u_{aL}\\
d_{aL}\end{pmatrix}$ & 3 & 2 & $1/6$ & $0$ & $+$\\
$u_{aR}$ & 3 & 1 & $2/3$ & $0$ & $+$\\
$d_{aR}$ & 3 & 1 & $-1/3$ & $0$ & $+$\\
$H=\begin{pmatrix}
H^+\\
H^0\end{pmatrix}$ & 1 & 2 & $1/2$ & 0 & $+$\\  
$\eta =\begin{pmatrix}
\eta^0\\
\eta^-\end{pmatrix}$ & 1 & 2 & $-1/2$ & $+1$ & $-$\\  
$\chi =\begin{pmatrix}
\chi^0\\
\chi^-\end{pmatrix}$ & 1 & 2 & $-1/2$ & $-1$ & $-$\\  
$\phi$ & 1 & 1 & 0 & 2 & $+$\\
\hline\hline 
\end{tabular}
\caption[]{\label{tab1} Field representation content of the model.}
\ec
\end{table}    

Further, $\nu_{1R}$ couples to leptons via the usual Higgs doublet $H$, while $\nu_{2,3R}$ couple to leptons via new Higgs doublets $\eta,\chi$ with charge $D=+1,-1$, respectively. $D$ charge is broken by a new Higgs singlet $\phi$ with charge $D=2$, which couples to $\nu_{2,3R}$, generating Majorana masses for $\nu_{2,3R}$ too. Notice that $\nu_{1R}$ can pick up a Majorana mass by itself, unsuppressed by any symmetry. It is noteworthy that $D$ charge is broken by two units due to $\langle \phi\rangle $ down to a residual dark parity $P_D=(-1)^D$. Particle representation content under the full gauge symmetry and residual dark parity is collected in Table \ref{tab1} for brevity. The dark parity conservation implies that $\eta,\chi$ have vanishing VEVs, while $H,\phi$ develop VEVs, $\langle H\rangle =(0,v/\sqrt{2})$ and $\langle \phi\rangle =\La/\sqrt{2}$, such that $\La\gg v=246$ GeV for consistency with the standard model. 

\begin{figure}[h]
\bc
\includegraphics[scale=1]{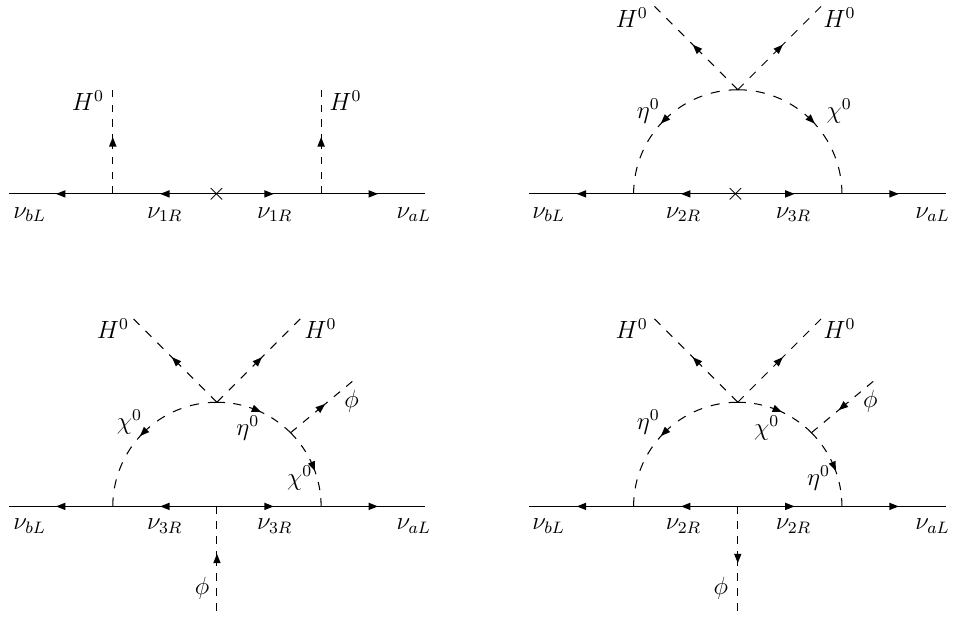}
\caption[]{\label{fig1} Scotoseesaw neutrino mass generation governed by dark gauge symmetry, where interchanging vertices of $\nu_{aL}$ and $\nu_{bL}$ leads to three new one-loop diagrams (unplotted).}
\ec
\end{figure}

It is further stressed that the neutrinos gain a suitable mass via a so-called scotoseesaw mechanism, as described by Feynman diagrams in Fig. \ref{fig1}. The tree-level diagram is a part of the canonical seesaw governed by $\nu_{1R}$, while the loop-level diagrams are a part of the scotogenic scheme set by $\nu_{2,3R}$. For comparison, the latter, which contains Majorana fermions in the loop, differs from that in \cite{dcz2}, which interprets Dirac fermions by contrast. Additionally, the present scotoseesaw recognized for three families naturally arises as a result of dark charge breaking, significantly overhauling the simplest models previously studied in \cite{valless,kubo}. Let us remind the reader that a recent approach \cite{valless5} based upon gauged $B-L$ that assigns $B-L=-4,-4,+5$ to $\nu_{1,2,3R}$ also hints at the existence of the scotoseesaw; furthermore, the scotogenic part may alternatively come from \cite{dongsg}.       

\section{\label{pot} Scalar potential}

We decompose the scalar potential into $V=V(H,\phi)+V(\eta,\chi,\mathrm{mix})$, where the first part includes the normal fields that induce gauge symmetry breaking,
\be V(H,\phi) = \mu^2_1 H^\dagger H + \mu^2_2 \phi^*\phi + \la_1(H^\dagger H)^2+\la_2 (\phi^*\phi)^2+\la_3 (H^\dagger H)(\phi^*\phi),\ee while the second part contains the dark fields and their couplings with normal fields,
\bea V(\eta,\chi,\mathrm{mix}) &=& \mu^2_3 \eta^\dagger \eta + \mu^2_4 \chi^\dagger \chi+\la_4(\eta^\dagger \eta)^2+\la_5(\chi^\dagger \chi)^2+\la_6(\eta^\dagger \eta)(\chi^\dagger \chi)+\la_7(\eta^\dagger \chi)(\chi^\dagger \eta)\crn
&& +(\la_8 \eta^\dagger \eta +\la_9 \chi^\dagger \chi)(H^\dagger H)+\la_{10} (\eta^\dagger H)(H^\dagger \eta)+\la_{11} (\chi^\dagger H)(H^\dagger \chi)\crn
&&+(\la_{12}\eta^\dagger \eta+\la_{13}\chi^\dagger \chi)(\phi^* \phi)+\left[\la_{14}(H\eta)(H\chi)+\mu_5 (\eta^\dagger \chi) \phi +H.c.\right].\eea 
Hereafter, $\la_{14}$ and $\mu_5$ are assumed to be real, since otherwise their phases can be absorbed by redefining appropriate scalar fields. The expected vacuum structure acquires $\mu^2_{1,2}<0$ and $\mu^2_{3,4}>0$. Additionally, the necessary condition for the potential bounded from below is $\la_{1,2,4,5}>0$, which is obtained by requiring $V>0$ for $H,\phi,\eta,\chi$ separately tending to infinity, respectively. Supplemental conditions would exist by requiring $V>0$ for a number of scalar fields simultaneously tending to infinity, which is skipped for brevity.  

Expanding $H^0=(v+S+i A)/\sqrt{2}$, $\phi=(\La +S'+iA')/\sqrt{2}$, $\eta^0=(R+i I)/\sqrt{2}$, and $\chi^0=(R'+i I')/\sqrt{2}$, we find the condition of potential minimum, 
\be v^2=\fr{2(\la_3 \mu^2_2-2\la_2 \mu^2_1)}{4\la_1\la_2-\la^2_3},\hs \La^2=\fr{2(\la_3 \mu^2_1-2\la_1 \mu^2_2)}{4\la_1\la_2-\la^2_3}.\ee Physical scalar fields relevant to the normal sector are identified as
\be H=\begin{pmatrix} G^+_W\\
\fr{1}{\sqrt{2}}(v+c_\varphi H_1+s_\varphi H_2+ i G_Z)
\end{pmatrix},\hs \phi=\fr{1}{\sqrt{2}}(\La -s_\varphi H_1+c_\varphi H_2 + i G_{Z'}),\ee  where $G^+_W\equiv H^+$, $G_Z\equiv A$, and $G_{Z'}\equiv A'$ are massless Goldstone bosons coupled to gauge bosons $W^+$, $Z$, and $Z'$, respectively. The standard model-like Higgs boson $H_1\equiv c_\varphi S - s_\varphi S'$ and the new Higgs boson $H_2\equiv s_\varphi S +c_\varphi S' $ are determined via a $S$-$S'$ mixing angle, \be t_{2\varphi}=\fr{\la_3 v \La}{\la_2 \La^2-\la_1 v^2}\simeq \fr{\la_3}{\la_2}\fr{v}{\La}\ll 1,\ee as well as obtaining respective masses,
\bea && m^2_{H_{1}}=\la_1 v^2+\la_2 \La^2 - \sqrt{(\la_1 v^2-\la_2 \La^2)^2+\la^2_3 v^2\La^2}\simeq \left(2\la_1-\fr{\la^2_3}{2\la_2}\right)v^2,\\
&&m^2_{H_{2}}=\la_1 v^2+\la_2 \La^2 + \sqrt{(\la_1 v^2-\la_2 \La^2)^2+\la^2_3 v^2\La^2}\simeq 2\la_2 \La^2.\eea      
        
Relevant to the dark scalar sector, we conveniently define $M^2_\eta\equiv \mu^2_3+\fr{\la_8}{2}v^2+\fr{\la_{12}}{2}\La^2$ and $M^2_\chi\equiv \mu^2_4+\fr{\la_9}{2}v^2+\fr{\la_{13}}{2}\La^2$. The neutral dark scalars $R,R'$ and $I,I'$ mix in each pair, such as 
\bea V &\supset& \fr 1 2 \begin{pmatrix} R & R' \end{pmatrix}
\begin{pmatrix} M^2_\eta & \fr{\mu_5 \La}{\sqrt{2}}+\fr{\la_{14}v^2}{2}\\
\fr{\mu_5 \La}{\sqrt{2}}+\fr{\la_{14}v^2}{2} & M^2_\chi \end{pmatrix}
\begin{pmatrix} R \\ R'\end{pmatrix}\crn
&& +\fr 1 2  \begin{pmatrix} I & I' \end{pmatrix}
\begin{pmatrix} M^2_\eta & \fr{\mu_5 \La}{\sqrt{2}}-\fr{\la_{14}v^2}{2}\\
\fr{\mu_5 \La}{\sqrt{2}}-\fr{\la_{14}v^2}{2} & M^2_\chi \end{pmatrix}
\begin{pmatrix} I \\ I'\end{pmatrix}. \eea Defining the mixing angles of the dark real and imaginary sectors, \be t_{2\theta_R}=\fr{\sqrt{2}\mu_5\La+\la_{14}v^2}{M^2_{\chi}-M^2_\eta},\hs t_{2\theta_I}=\fr{\sqrt{2}\mu_5\La-\la_{14}v^2}{M^2_{\chi}-M^2_\eta},\ee we obtain physical fields,
\bea && R_1=c_{\theta_R} R -s_{\theta_R} R',\hs R_2=s_{\theta_R} R +c_{\theta_R} R',\\
&& I_1=c_{\theta_I} I -s_{\theta_I} I',\hs I_2=s_{\theta_I} I +c_{\theta_I} I', \eea with respective masses,
\bea && m^2_{R_{1}}\simeq M^2_\eta + \fr{\fr 1 4 (\sqrt{2}\mu_5\La+\la_{14} v^2)^2}{M^2_\eta - M^2_\chi},\hs m^2_{R_{2}}\simeq M^2_\chi - \fr{\fr 1 4 (\sqrt{2}\mu_5\La+\la_{14} v^2)^2}{M^2_\eta - M^2_\chi},\\
&& m^2_{I_{1}}\simeq M^2_\eta + \fr{\fr 1 4 (\sqrt{2}\mu_5\La-\la_{14} v^2)^2}{M^2_\eta - M^2_\chi},\hs m^2_{I_{2}}\simeq M^2_\chi - \fr{\fr 1 4 (\sqrt{2}\mu_5\La-\la_{14} v^2)^2}{M^2_\eta - M^2_\chi},\eea where the approximations assume $\mu_5\La\sim \la_{14}v^2\ll M^2_{\eta,\chi}\sim \La^2$, thus $\mu_5\ll \La$. It is clear that $R_1,I_1$, as well as $R_2,I_2$, are now separated in mass, namely $(m^2_{R_i}-m^2_{I_i})/(m^2_{R_i}+m^2_{I_i})\sim \la^2_{14}(v/\La)^4\ll 1$, for $i=1,2$. It is stressed that this mass splitting is of the same magnitude as the mixing angles squared, $\theta^2_{R,I}\sim \la^2_{14}(v/\La)^4\ll 1$.

Similarly, the charged dark scalars $\eta^-$ and $\chi^-$ mix by themselves, such as
\bea V\supset \begin{pmatrix} \eta^+ & \chi^+\end{pmatrix}\begin{pmatrix}
M^2_\eta +\fr{\la_{10}}{2}v^2 & \fr{\mu_5\La}{\sqrt{2}}\\
\fr{\mu_5\La}{\sqrt{2}} & M^2_\chi+\fr{\la_{11}}{2}v^2 \end{pmatrix}\begin{pmatrix} \eta^-\\
\chi^-\end{pmatrix}.\eea Diagonalizing the relevant mass matrix, we obtain physical fields, \be C^-_1=c_{\theta_C} \eta^- -s_{\theta_C}\chi^-,\hs C^-_2=s_{\theta_C} \eta^-+c_{\theta_C}\chi^-,\ee where the mixing angle of the charged dark scalar sector obeys \be t_{2\theta_C}=\fr{\sqrt{2}\mu_5\La}{M^2_\chi-M^2_\eta+\fr 1 2 (\la_{11}-\la_{10})v^2}.\ee The masses of physical charged dark fields are approximated by
\bea m^2_{C_1} &\simeq& M^2_\eta +\fr{\la_{10}}{2}v^2+\fr{\fr 1 2 \mu^2_5 \La^2}{M^2_\eta -M^2_\chi +\fr 1 2 (\la_{10}-\la_{11})v^2},\\ 
m^2_{C_2} &\simeq& M^2_\chi+\fr{\la_{11}}{2}v^2-\fr{\fr 1 2 \mu^2_5 \La^2}{M^2_\eta -M^2_\chi +\fr 1 2 (\la_{10}-\la_{11})v^2}.\eea Because of $\mu_5\La\sim \la_{14} v^2\ll M^2_{\eta,\chi}\sim \La^2$, the charged dark scalar mixing is small, similar to neutral dark sectors, i.e. $\theta_C\sim \theta_{R,I}$.

\section{\label{neum} Neutrino mass} 

Every charged lepton and quark gain an appropriate mass through interacting with the Higgs field $H$ identical to the standard model. Concerning left-handed and right-handed neutrinos, their Yukawa Lagrangian is given by 
\bea \mathcal{L} &\supset& h_{a1}\bar{l}_{aL}\tilde{H}\nu_{1R}+h_{a2}\bar{l}_{aL}\eta \nu_{2R}+h_{a3}\bar{l}_{aL}\chi \nu_{3R}\crn
&&-\fr 1 2 M_{1} \bar{\nu}^c_{1R}\nu_{1R}+\fr 1 2 f_{2} \bar{\nu}^c_{2R}\nu_{2R}\phi +\fr 1 2 f_{3} \bar{\nu}^c_{3R}\nu_{3R}\phi^* -M_{23}\bar{\nu}^c_{2R} \nu_{3R}+H.c.\label{eqadd121}\eea 
When $H,\phi$ develop VEVs, we obtain a tree-level mass Lagrangian,  
\be \mathcal{L}\supset -\fr 1 2 \begin{pmatrix} \bar{\nu}_{aL} & \bar{\nu}^c_{1R}\end{pmatrix}\begin{pmatrix} 0 & m_{a1}\\
m_{b1} & M_1 \end{pmatrix}\begin{pmatrix} \nu^c_{bL} \\
\nu_{1R}\end{pmatrix} -\fr 1 2 \begin{pmatrix}\bar{\nu}^c_{2R} & \bar{\nu}^c_{3R}\end{pmatrix}
\begin{pmatrix} M_{2} & M_{23}\\
M_{23} & M_{3}\end{pmatrix}\begin{pmatrix}
\nu_{2R}\\
\nu_{3R}\end{pmatrix}+H.c.,\label{dtttn}\ee where $b=1,2,3$ is a family index, as $a$ is. Additionally, $m_{a1} = -h_{a1}v/\sqrt{2}$ is a Dirac mass, while $M_{2} = -f_2\La/\sqrt{2}$ and $M_{3} = -f_3\La/\sqrt{2}$ are Majorana masses, as $M_1$ and $M_{23}$ are. 

Without loss of generality, we assume $M_{23}=0$, i.e. neither mixing between $\nu_{2R}$ and $\nu_{3R}$ nor the corresponding loop diagram (upper right) in Fig. \ref{fig1} exist. Diagonalizing the mass matrix of $(\nu_{L},\nu_{1R})$ in (\ref{dtttn}), which has a seesaw form due to $v\ll M_1$, we get 
\be \mathcal{L}\supset -\fr 1 2 (m_\nu)^{\mathrm{tree}}_{ab}\bar{\nu}_{aL}\nu^c_{bL}-\fr 1 2 M_1 \bar{\nu}^c_{1R}\nu_{1R} -\fr 1 2 M_2 \bar{\nu}^c_{2R}\nu_{2R}-\fr 1 2 M_3 \bar{\nu}^c_{3R}\nu_{3R}+H.c.,\ee where the first term,
\be (m_\nu)^{\mathrm{tree}}_{ab}\simeq -\fr{m_{a1}m_{b1}}{M_1},\label{dtttn1}\ee is the seesaw-induced neutrino mass, identical to the contribution of the tree-level diagram (upper left) in Fig. \ref{fig1}, while $\nu_{1,2,3R}$ are apparently decoupled, behaving as physical Majorana fields with masses $M_{1,2,3}$, respectively.  

The tree-level mass matrix (\ref{dtttn1}) gives only a nonzero (mass) eigenvalue according to mass eigenstate $\sim m_{a1}\nu_{aL}$, which is inappropriate for the experiment \cite{pdg}. Interestingly, the tree-level vanished masses, or exactly the mass matrix of ($\nu_L,\nu_{1R}$) in (\ref{dtttn}), receive radiative corrections by dark fields $(\eta,\chi,\nu_{2,3R})$, as in the loop diagrams (lower left and right) in Fig.~\ref{fig1}. On a mass basis, the loop diagrams are commonly contributed by one of the $\nu_{2,3R}$ fermions and one of the $R_{1,2},I_{1,2}$ scalars, determined by the Lagrangian,
\bea \mathcal{L} &\supset& \fr{h_{a2}}{\sqrt{2}}\bar{\nu}_{aL}\nu_{2R}(c_R R_1+s_R R_2+i c_I I_1+i s_I I_2)\crn
&&+\fr{h_{a3}}{\sqrt{2}}\bar{\nu}_{aL}\nu_{3R}(-s_R R_1+c_R R_2-i s_I I_1+i c_I I_2)+H.c.\label{eqadd11}\eea plus kinetic and mass terms of the relevant physical fields, where we have defined $c_R\equiv c_{\theta_R}$ and $s_R\equiv s_{\theta_R}$ for short. Therefore, the loop-induced neutrino masses are computed as 
\bea (m_\nu)^{\mathrm{rad}}_{ab}&=&\fr{h_{a2}h_{b2} M_2}{32\pi^2}\left(\fr{c^2_R m^2_{R_1} \ln \fr{M^2_2}{m^2_{R_1}}}{M^2_2-m^2_{R_1}}- \fr{c^2_I m^2_{I_1} \ln \fr{M^2_2}{m^2_{I_1}}}{M^2_2-m^2_{I_1}}+ \fr{s^2_R m^2_{R_2} \ln \fr{M^2_2}{m^2_{R_2}}}{M^2_2-m^2_{R_2}}- \fr{s^2_I m^2_{I_2} \ln \fr{M^2_2}{m^2_{I_2}}}{M^2_2-m^2_{I_2}}\right)\crn
&&+\nu_{3R}\mathrm{-contribution},\label{raddtttn}\eea where the $\nu_{3R}$--contribution is translated from that of $\nu_{2R}$ (previous term) by replacing $h_{a(b)2}\to h_{a(b)3}$, $M_2\to M_3$, $c_{R,I}\to -s_{R,I}$, and $s_{R,I}\to c_{R,I}$. Notice that the divergences are manifestly cancelled out due to $c^2_R-c^2_I+s^2_R-s^2_I=0$ for each $\nu_{2,3R}$ contribution. 

With the aid of dark scalar mass splittings and mixing angles, i.e. $(m^2_{R_i}-m^2_{I_i})/(m^2_{R_i}+m^2_{I_i})\sim \theta^2_{R,I}\sim \la^2_{14}(v/\La)^4$ for $i=1,2$, we estimate the magnitude of radiative masses,
$(m_\nu)^{\mathrm{rad}}\sim 0.1 (M_{2,3}/\mathrm{TeV})(h\la_{14}/10^{-3})^2\ \mathrm{eV}$, assuming $v/\La\sim 10^{-1}$, where $h\sim h_{a(b)k}$. Comparing with the measured value $m_\nu\sim 0.1$ eV implies $M_{2,3}\sim $ TeV and $h\la_{14}\sim 10^{-3}$, thus $h,\la_{14}$ are not too small. The last conclusion is due to a suppression in dark scalar mass splittings and mixing angles as $(v/\La)^4$, which is more stronger than that of the conventional scotogenic model \cite{ma}. On the other hand, the tree-level neutrino mass in (\ref{dtttn1}) obeys $(m_\nu)^{\mathrm{tree}}\sim 0.1$ eV, indicating that either $M_1$ is close to the GUT scale for $h_{a1}\sim h\sim 0.1$, or alternatively $h_{a1}$ is close to the Yukawa coupling of electrons for $M_1\sim M_{2,3}\sim$ TeV. The first case is possible since $M_1$ is not suppressed by dark symmetry, while the second case is available too but it requires a hierarchy $h_{a1}\ll h_{ak}$ like those of the Yukawa couplings of charged leptons. In summary, the scotoseesaw neutrino mass described in Fig. \ref{fig1} is given by  \be (m_\nu)_{ab} = (m_\nu)^{\mathrm{tree}}_{ab}+(m_\nu)^{\mathrm{rad}}_{ab}.\label{adt1}\ee 

The full neutrino mass matrix (\ref{adt1}) can be rewritten as 
\be m_\nu = h\Delta h^T,\ee where $h=(h_{an})_{3\times 3}$ is the $3\times 3$ Yukawa coupling matrix that  arranges the existing couplings $h_{a1}$, $h_{a2}$, and $h_{a3}$ in the columns $n=1,2,3$, respectively. Comparing $(m_\nu)_{ab}=h_{a n}h_{b n'}\Delta_{nn'}$ ($n'=1,2,3$, like $n$) with (\ref{dtttn1}) and (\ref{raddtttn}), we obtain $\Delta=\mathrm{diag}(\Delta_1,\Delta_2,\Delta_3)$ to be a $3\times 3$ diagonal matrix, with the elements $\Delta_{1,2,3}$ collected as
\bea \Delta_1 &=& -\fr{v^2}{2M_1},\\
\Delta_2 &=&\fr{M_2}{32\pi^2}\left(\fr{c^2_R m^2_{R_1} \ln \fr{M^2_2}{m^2_{R_1}}}{M^2_2-m^2_{R_1}}- \fr{c^2_I m^2_{I_1} \ln \fr{M^2_2}{m^2_{I_1}}}{M^2_2-m^2_{I_1}}+ \fr{s^2_R m^2_{R_2} \ln \fr{M^2_2}{m^2_{R_2}}}{M^2_2-m^2_{R_2}}- \fr{s^2_I m^2_{I_2} \ln \fr{M^2_2}{m^2_{I_2}}}{M^2_2-m^2_{I_2}}\right),\\
\Delta_3&=&\Delta_2(M_2\to M_3;c_{R,I}\to -s_{R,I}; s_{R,I}\to c_{R,I}). \eea Therefore, we can use the Casas-Ibarra parametrization \cite{cipara} to estimate the Yukawa coupling matrix $h$, such as
\be h = U_\nu D_{\sqrt{m_\nu}} T D_{\sqrt{\Delta^{-1}}},\ee where $T$ is an $3\times 3$ arbitrary matrix that obeys the orthogonal condition, i.e. $TT^T=1$, $U_\nu$ is the Pontecorvo-Maki-Nakawaga-Sakata (PMNS) matrix that diagonalizes the neutrino mass matrix, i.e. $U^T_\nu m_\nu U_\nu =\mathrm{diag}(m_1,m_2,m_3)$, $D_{\sqrt{m_\nu}}=\mathrm{diag}(\sqrt{m}_1,\sqrt{m}_2,\sqrt{m}_3)$, and $D_{\sqrt{\Delta^{-1}}}=\mathrm{diag}(\sqrt{\Delta^{-1}_1},\sqrt{\Delta^{-1}_2},\sqrt{\Delta^{-1}_3})$. Notice that we can choose bases and parameters as appropriate so that $m_{1,2,3}$ and $\Delta_{1,2,3}$ are all real and positive.

If the mixing between $\nu_{2R}$ and $\nu_{3R}$ is included, the above result of radiative neutrino masses (\ref{raddtttn}) is easily generalized by changing $h_{a(b)k}\to h_{a(b)k}U_{kl}$ for $k,l=2,3$, where $U$ relates $\nu_{2,3R}$ to mass eigenstates $\nu'_{2,3R}$, such as $\nu_{kR}=U_{kl}\nu'_{lR}$, with mass eigenvalues $M'_{2,3}$ instead of $M_{2,3}$. Indeed, we diagonalize the relevant mass matrix in (\ref{dtttn}) to identify the physical states $\nu'_{2R}=c_\xi \nu_{2R}-s_\xi \nu_{3R}$ and $\nu'_{3R}=s_\xi \nu_{2R}+c_\xi \nu_{3R}$ with respective masses $M'_{2,3}=\fr 1 2 [M_2+M_3\mp \sqrt{(M_2-M_3)^2+4M_{23}}]$, where $t_{2\xi}=2 M_{23}/(M_3-M_2)$, thus \be U=\begin{pmatrix} c_\xi & s_\xi \\
-s_\xi & c_\xi \end{pmatrix}.\ee The Lagrangian (\ref{eqadd11}) becomes \bea \mathcal{L} &\supset& \fr{1}{\sqrt{2}}\bar{\nu}_{aL}\nu'_{lR}\left[(h'_{al}c_R-h''_{al}s_R) R_1 + (h'_{al} s_R + h''_{al} c_R)R_2\right.\crn
&&\left.+i(h'_{al}c_I-h''_{al}s_I) I_1 + i (h'_{al} s_I + h''_{al} c_I)I_2\right]+H.c.,\eea where we define $h'_{al} = h_{a2}U_{2l}$ and $h''_{al} = h_{a3}U_{3l}$, for brevity. It is straightforward to compute radiative neutrino masses in this case, such as 
\bea (m_\nu)^{\mathrm{rad}}_{ab}&=&\fr{M'_l}{32\pi^2}\left[(h'_{al}c_R-h''_{al}s_R)(h'_{bl}c_R-h''_{bl}s_R)\fr{m^2_{R_1} \ln \fr{M'^2_l}{m^2_{R_1}}}{M'^2_l-m^2_{R_1}} \right.\crn
&&\left. - (h'_{al}c_I-h''_{al}s_I)(h'_{bl}c_I-h''_{bl}s_I)\fr{ m^2_{I_1} \ln \fr{M'^2_l}{m^2_{I_1}}}{M'^2_l-m^2_{I_1}}\right.\crn
&&\left.+ (h'_{al} s_R + h''_{al} c_R)(h'_{bl} s_R + h''_{bl} c_R)\fr{ m^2_{R_2} \ln \fr{M'^2_l}{m^2_{R_2}}}{M'^2_l-m^2_{R_2}}\right.\crn
&&\left. - (h'_{al} s_I + h''_{al} c_I)(h'_{bl} s_I + h''_{bl} c_I)\fr{ m^2_{I_2} \ln \fr{M'^2_l}{m^2_{I_2}}}{M'^2_l-m^2_{I_2}}\right].\label{eqadd12}\eea It is stressed that the mixing $\nu_{2R}$-$\nu_{3R}$ simply modifies their couplings---those of the dark right-handed neutrinos---to the usual neutrinos and the dark scalars by a rotation matrix $U$ with angle $\xi$. The result in (\ref{eqadd12}) matches the original one in (\ref{raddtttn}) in the limit $\xi\to 0$, or equivalently shifting $h'_{a(b)l}\to h_{a(b)2}$, $h''_{a(b)l}\to h_{a(b)3}$, and $M'_l\to M_l$, as appropriate. On the other hand, the full neutrino mass matrix is $m_\nu=m^{\mathrm{tree}}_\nu+m^{\mathrm{rad}}_\nu =h\Delta h^T$, which combines (\ref{eqadd12}) instead of (\ref{raddtttn}), in which $\Delta$ is now not a diagonal matrix, as it possesses a nonzero off-diagonal element $\Delta_{23}=\Delta_{32}\neq 0$, different from the previous case. We can derive the Yukawa coupling matrix $h$ by diagonalizing $\Delta$ similarly to $m_\nu$, i.e. $O\Delta O^T=\mathrm{diag}(\Delta_1,\Delta_2,\Delta_3)$, which yields $h = U_\nu D_{\sqrt{m_\nu}} T D_{\sqrt{\Delta^{-1}}}O$ \cite{aapara}. Note that $\Delta_{23}=0$ and $O=1$ in the limit $\xi\to 0$. It is clear that the physics does not significantly change when including the $\nu_{2R}$-$\nu_{3R}$ mixing effect. Hence, we would take $\xi=0$, i.e. the result in (\ref{raddtttn}) for neutrino mass, into account, as also aforementioned.       

It is noted that the normal Higgs fields $H_{1,2}$ and their relevant Goldstone ($G_{Z,Z'}$) and gauge ($Z,Z'$) bosons may mediate new one-loop diagrams, contributing to neutrino masses. Here, the inclusion of the Goldstone diagrams cancels the divergent part of the diagrams mediated by $H_{1,2}$. However, we show in Appendix \ref{appa} that such contributions are radically smaller than the scotoseesaw one, which would be skipped.           
    
\section{\label{darkm} Dark matter}

The model contains two kinds of dark matter, dark Majorana fermion ($\nu_{2,3R}$) and dark doublet scalar ($R_{1,2},I_{1,2}$), communicated with normal matter via the $U(1)_D$ gauge boson $Z'$ and usual/new Higgs portals $H_{1,2}$. The kinetic mixing effect \cite{kimix} between $U(1)_D$ and $U(1)_Y$ is too small and is not signified in this work. The dark matter candidate---the lightest of the dark fields by an appropriate parameter choice---is studied in order.  

\subsection{Majorana dark matter}

\begin{figure}[h]
\bc
\includegraphics[scale=1]{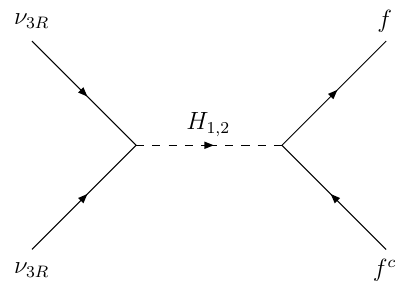}\hspace{1cm} 
\includegraphics[scale=1]{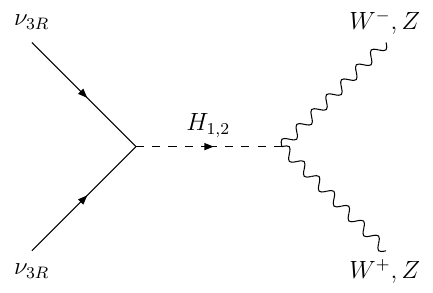}\\ 
\includegraphics[scale=1]{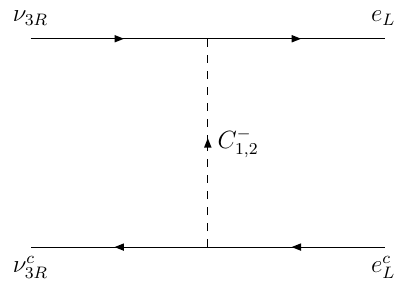}\hspace{1cm}
\includegraphics[scale=1]{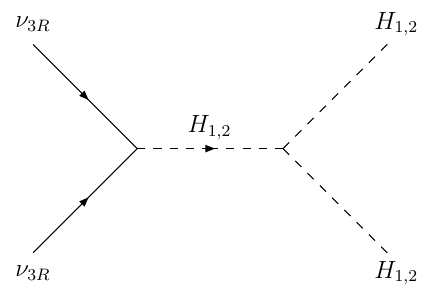}\\
\includegraphics[scale=1]{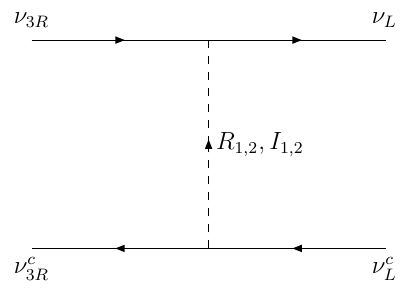}\hspace{1cm}
\includegraphics[scale=1]{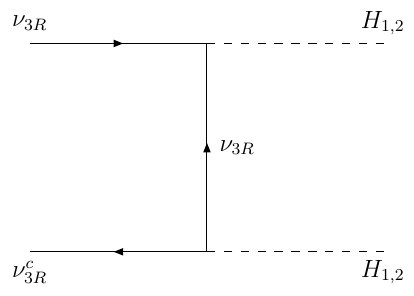}
\caption[]{\label{fig2} Annihilation of Majorana dark matter to normal matter that sets the relic abundance. It is noted that each $t$-channel diagram has a corresponding $u$-channel diagram due to the Majorana nature of incoming dark matter, which is not displayed (see, e.g., \cite{fthem,annvector}).}
\ec
\end{figure}

\begin{figure}[h]
\bc
\includegraphics[scale=1]{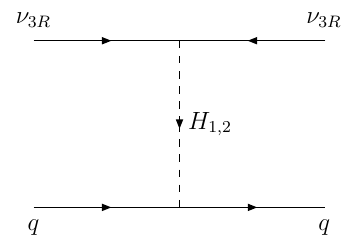}
\caption[]{\label{fig3} Scattering of Majorana dark matter with nuclei in direct detection.}
\ec
\end{figure}

Assuming $\nu_{3R}$ to be the lightest of the dark fields, it is absolutely stabilized by the residual dark parity $P_D$ responsible for dark matter. In the early universe, $\nu_{3R}$'s potentially annihilate to a pair of $Z'$---given that $Z'$ is lighter than $\nu_{3R}$---via $t,u$-channel diagrams as well as to usual fermions ($ff^c$) and gauge/Higgs bosons ($W^-W^+$, $ZZ$, and $H_{1,2}H_{1,2}$) via $s$-channel $H_{1,2}$ portals, mostly due to a mixing between the standard model Higgs field ($H$) and new Higgs field ($\phi$), and/or via $t,u$-channel diagrams mediated by appropriate dark fields.\footnote{Notice that the dark matter annihilation to $H_1 H_1$ will be insignificant if the $H,\phi$ coupling (i.e., $\la_3$) is not large, while the dark matter annihilation to $H_2 H_2$ will be kinematically suppressed if $H_2$ is heavier than the dark matter, which must be taken into account in this work.} The usual Higgs $H_1$ portal negligibly contributes to the relic density since $\nu_{3R}$ is heavy at TeV, necessarily set by $H_2$ resonance. Also, the annihilation to $Z'$ by $t,u$-channels is suppressed similarly to the $H_1$ portal, and not significantly compared to the $H_2$ portal (cf. \cite{dpsearch}). That said, the $\nu_{3R}$ annihilation that sets the relic density is governed by the $H_2$ portal, as described by the Feynman diagrams in Fig.~\ref{fig2}. Here, for the verification of the $H_1$ contribution, those diagrams with the $H_1$ portal are also included, whereas the annihilation to $Z'$ is neglected by assuming $m_{Z'}>M_3$. Similarly, the nonresonant annihilations to leptons and Higgs fields via $t,u$-channel dark-field exchanges are small but are included in Fig.~\ref{fig2} for convenience in the comparison. Although the $H_1$ portal does not contribute to the relic density, it would set the spin-independent (SI) cross section, presenting scattering of $\nu_{3R}$ with nuclei in the direct detection experiment, as determined by the Feynman diagram in Fig. \ref{fig3}. This is because the $H_2$ contribution to the effective interaction $(\bar{\nu}^c_{3R}\nu_{3R})(\bar{q}q)$ is suppressed by $m^2_{H_1}/m^2_{H_2}\ll 1$, as compared to that of $H_1$.  

The annihilation cross section of $\nu_{3R}$'s to quarks is given via $H_{1,2}$ exchanges, as in the top-left diagram in Fig. \ref{fig2}, evaluated as 
\bea \langle \sigma v_{\mathrm{rel}}\rangle_{\nu_{3R}\nu_{3R}\to \mathrm{quarks}} &=& \sum_q \fr{s^2_\varphi c^2_\varphi m^2_q M^4_3}{8\pi v^2\La^2 }\left(\fr{1}{4 M^2_3-m^2_{H_2}}-\fr{1}{4M^2_3-m^2_{H_1}}\right)^2 \left(1-\fr{m^2_q}{M^2_3}\right)^{\fr 3 2}\crn
&\simeq& \fr{s^2_\varphi m^2_t M^4_3}{8\pi v^2 \La^2}\left(\fr{1}{4M^2_3-m^2_{H_2}}-\fr{1}{4M^2_3-m^2_{H_1}}\right)^2 \left(1-\fr{m^2_t}{M^2_3}\right)^{\fr 3 2},\label{themq} \eea where $v_{\mathrm{rel}}$ is dark matter relative velocity, and the biggest contribution comes from the annihilation to the top quark, as enhanced by its Yukawa coupling. Additionally, the annihilation cross section of $\nu_{3R}$'s to leptons is contributed by $s$-channel $H_{1,2}$ exchanges, as in the top-left diagram, and $t,u$-channel dark-field exchanges, as in the middle-left and bottom-left diagrams in Fig. \ref{fig2}. The $s$-channel contribution is identical to the above result by replacing $q\to \nu,e$ (i.e., leptons), for which this contribution is very small compared to the annihilation to top quarks, since $m^2_{\nu,e}\ll m^2_t$, which is safely suppressed. The $t,u$-channel contributions by the dark fields are given by\footnote{Ref. \cite{fthem} gives a general formula of the annihilation cross section, which is necessarily applied to an interesting case for dark matter and mediator masses to be low, at a weak scale, as further analyzed in \cite{annvector}, opposite to our current case, where all dark matter and mediator masses would be at TeV.} 
\bea \langle \sigma v_{\mathrm{rel}}\rangle_{\nu_{3R}\nu_{3R}\to \mathrm{leptons}} &=&\sum_{\nu=\nu_1,\nu_2,\nu_3}\fr{|h_{\nu 3}h^*_{\nu 3}|^2 M_3^2}{32\pi (M^2_3+m^2_{\chi^0})^2}\left(1-\fr{m^2_\nu}{M^2_3}\right)^{1/2} \crn
&&+ \sum_{e=e,\mu,\tau}\fr{|h_{e 3}h^*_{e 3}|^2 M_3^2}{32\pi (M^2_3+m^2_{\chi^-})^2}\left(1-\fr{m^2_e}{M^2_3}\right)^{1/2}\crn
&\simeq & \sum_{a=1,2,3}\fr{|h_{a3}h^*_{a3}|^2 M^2_3}{32\pi}\left[\fr{1}{(M^2_3+m^2_{\chi^0})^2}+\fr{1}{(M^2_3+m^2_{\chi^-})^2}\right].\label{theml} \eea Here, we have neglected the small mixing effects of dark scalars due to $\theta_C\sim \theta_{R,I}\ll 1$ and taken $m_{\chi^-}\simeq m_{C^-_2}$ and $m_{\chi^{0}}\simeq m_{R_2,I_2}$. Additionally, the last approximation comes from the fact that the lepton masses are substantially smaller than the dark matter mass, i.e. $m_{\nu,e}\ll M_3$. It is also noted that $h_{\nu 3}=\sum_a h_{a3}(V^*_{\nu L})_{a \nu}$ and $h_{e3}=\sum_a h_{a3}(V^*_{eL})_{a e}$, where $V_{\nu L}$ is the left-handed neutrino mixing matrix that diagonalizes the relevant neutrino mass matrix, i.e. $V^\dagger_{\nu L} m_\nu V^*_{\nu L}=\mathrm{diag}(m_{\nu_1},m_{\nu_2},m_{\nu_3})$, while $V_{e L}$ is the left-handed charged-lepton mixing matrix, which along with the right-handed charged-lepton mixing matrix $V_{eR}$ bring them to mass eigenstates, i.e. $V^\dagger_{eL}m_e V_{eR}=\mathrm{diag}(m_e,m_\mu,m_\tau)$. Two remarks are in order 
\ben
\item Because the dark matter $\nu_{3R}$ has a mass at the TeV scale like those of $H_2$, $\chi^0$, and $\chi^-$, which all are proportional to $\La$, the cross section (\ref{theml}) is similar in size to the cross section (\ref{themq}) for $M_3\neq \fr 1 2 m_{H_2}$, where we note that $s_\varphi\sim 10^{-2}$ (see below) and $h_{a3}\sim 0.1$ as given in the section on neutrino mass. When the dark matter relic density is set by $H_2$ resonance, i.e. $M_3=\fr 1 2 m_{H_2}$, the cross section (\ref{themq}) is strongly enhanced, while the $t,u$-channel contribution (\ref{theml}) is negligible. 
\item Besides the separate contributions of the $s$-channel and the $t,u$-channels to the dark matter annihilation to leptons, as given above, there is an interference term between two kinds of these contributions, because of $|M(\nu_{3R}\nu_{3R}\to\mathrm{leptons})|^2=|M_s+M_{t,u}|^2=|M_s|^2+ |M_{t,u}|^2+2\Re(M^*_sM_{t,u})$. The contribution of the interference term, i.e. $2\Re(M^*_sM_{t,u})$, is small too, since $M_s$ is very small, compared to that for annihilating to top quarks, thus $M_s\ll M_{t,u}$ for $M_3\neq \fr 1 2 m_{H_2}$. Otherwise, at the $H_2$ resonance, both $M_s$ and $M_{t,u}$ are insignificant. Hence, the interference term is safely omitted.        
\een

On the other hand, the annihilation cross section of $\nu_{3R}$'s to gauge bosons is given by the top-right diagram in Fig. \ref{fig2}, computed as \bea \langle \sigma v_{\mathrm{rel}}\rangle_{\nu_{3R}\nu_{3R}\to \mathrm{vectors}} &=& \sum_{V=W,Z}\fr{\kappa_V c^2_\varphi s^2_\varphi M^6_3}{ 4 \pi v^2 \La^2 }\left(\fr{1}{4M^2_3-m^2_{H_2}}-\fr{1}{4M^2_3-m^2_{H_1}} \right)^2\crn
&&\times \left(1-\fr{m^2_V}{M^2_3}\right)^{\fr 1 2}\left(1-\fr{m^2_V}{M^2_3}+\fr 3 4 \fr{m^4_V}{M^4_3}\right)\crn
&\simeq& \fr{3s^2_\varphi M^6_3}{8\pi v^2\La^2 }\left(\fr{1}{4M^2_3-m^2_{H_2}}-\fr{1}{4M^2_3-m^2_{H_1}} \right)^2, \eea where the statistical factor $\kappa_V$ is accounted for the creation of identical particles, i.e. $\kappa_V=1$ and $\fr 1 2$ for $V=W$ and $Z$, respectively.\footnote{The small contributions suppressed by $m^2_{W,Z}/M^2_3$ have been omitted, but that of $H_1$ given in the denominator of the propagator---which is small too---is retained for brevity (without confusion).} This result implies that the annihilation to $W,Z$ is more enhanced than the annihilation to top quarks for dark matter mass $M_3$ to be beyond the weak scale (cf. \cite{annvector}). The annihilation cross section of $\nu_{3R}$'s to Higgs bosons is described by the middle-right and bottom-right diagrams in Fig. \ref{fig2}, evaluated as \be \langle \sigma v_{\mathrm{rel}}\rangle_{\nu_{3R}\nu_{3R}\to \mathrm{scalars}} \simeq \langle \sigma v_{\mathrm{rel}}\rangle_{\nu_{3R}\nu_{3R}\to H_1 H_1} + \langle \sigma v_{\mathrm{rel}}\rangle_{\nu_{3R}\nu_{3R}\to H_2 H_2}, \ee where
\bea \langle \sigma v_{\mathrm{rel}}\rangle_{\nu_{3R}\nu_{3R}\to H_1 H_1}&\simeq & \fr{\la^2_3 M^2_3}{128\pi}\left[\fr{3m^2_{H_1}}{m^2_{H_2}(4M^2_3-m^2_{H_1})}-\fr{1}{4M^2_3-m^2_{H_2}}\right]^2\left(1-\fr{m^2_{H_1}}{M^2_3}\right)^{\fr 1 2},\\
\langle \sigma v_{\mathrm{rel}}\rangle_{\nu_{3R}\nu_{3R}\to H_2 H_2}&\simeq& \fr{1}{128\pi}\left[\fr{9M^2_3 m^4_{H_2}}{\La^4 (4M^2_3-m^2_{H_2})^2}+\fr{64\langle v^2\rangle M^6_3(M^2_3-m^2_{H_2})^2}{3\La^4 (2M^2_3-m^2_{H_2})^4}\right.\crn
&&\left. -\fr{8\langle v^2\rangle M^4_3 m^2_{H_2}(M^2_3-m^2_{H_2})}{\La^4(4M^2_3-m^2_{H_2})(2M^2_3-m^2_{H_2})^2}\right]\left(1-\fr{m^2_{H_2}}{M^2_3}\right)^{\fr 1 2}\Theta(M_3-m_{H_2}),\eea where $\Theta(\cdots)$ is the Heaviside step function. It is noted that the annihilation $\nu_{3R}\nu_{3R}\to H_1 H_2$ is more suppressed than the two mentioned processes, which would be neglected. Additionally, the $t,u$-channel diagrams that annihilate to $H_1 H_1$ negligibly contribute to the annihilation process $\nu_{3R}\nu_{3R}\to H_1 H_1$, which proceeds only via the $s$-channel diagrams by $H_{1,2}$ exchanges. On the other hand, the $s$-channel diagram by $H_1$ exchange negligibly contributes to the annihilation process $\nu_{3R}\nu_{3R}\to H_2 H_2$. Further, the $t,u$-channel contributions to $\nu_{3R}\nu_{3R}\to H_2 H_2$ are suppressed by $\langle v^2\rangle =3/2x_F$, where $x_F=M_3/T_F\simeq 25$ is given at freeze-out temperature.\footnote{The quantity $v$ existing only in the thermal average $\langle v^2\rangle$ is dark matter velocity, i.e. $v=\fr 1 2 v_{\mathrm{rel}}$, which should not be confused with the weak scale used throughout.} Last, but not least, the annihilation cross section to $H_1H_1$ is comparable to that to $W,Z$, given that $\la_3\sim s_\varphi (M_3/v)\sim 0.1$. The annihilation cross section to $H_2H_2$ is the biggest, if kinematically allowed/opened, as the process is not suppressed by small couplings, say $s_\varphi$, $h_{3a}$, or $\la_3$, as in the previous processes. Indeed, it is governed by the Yukawa coupling $-\fr{f_3}{\sqrt{2}}=\fr{M_3}{\La}$ and is thus enhanced if $M_3$ is raised (for fixed $\La$) before being prevented by the unitarity condition.        

\begin{figure}[h]
\bc
\includegraphics[scale=1]{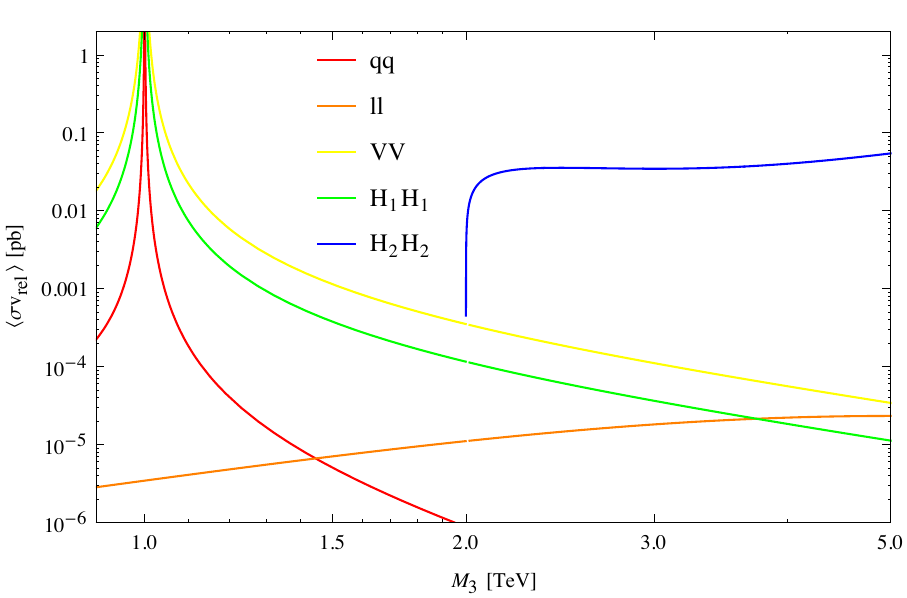}
\caption[]{\label{fig4} Majorana dark matter annihilation cross sections to several channels plotted as functions of dark matter mass for a comparison.}
\ec
\end{figure}

\begin{figure}[h]
\bc
\includegraphics[scale=1]{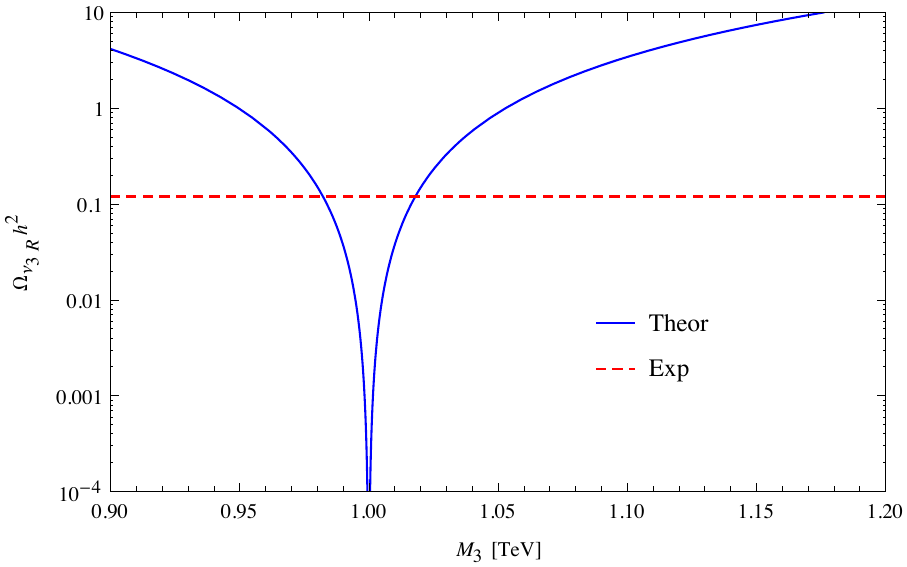}
\caption[]{\label{fig5} Relic density of Majorana dark matter plotted as a function of its mass, where the dashed line is the measured value $\Om_{\mathrm{DM}} h^2 \simeq 0.12$ \cite{pdg}.}
\ec
\end{figure}

The total dark matter annihilation cross section is summed over the fermionic and bosonic channels, such as 
\bea \langle \sigma v_{\mathrm{rel}}\rangle_{\nu_{3R}} &=& \langle \sigma v_{\mathrm{rel}}\rangle_{\nu_{3R}\nu_{3R}\to \mathrm{quarks}}+\langle \sigma v_{\mathrm{rel}}\rangle_{\nu_{3R}\nu_{3R}\to \mathrm{leptons}}\crn
&&+\langle \sigma v_{\mathrm{rel}}\rangle_{\nu_{3R}\nu_{3R}\to \mathrm{vectors}}+\langle \sigma v_{\mathrm{rel}}\rangle_{\nu_{3R}\nu_{3R}\to \mathrm{scalars}}. \eea The usual-new Higgs mixing leads to a shift in the standard model Higgs couplings, which are well measured at present, and the projected HL-LHC would make stringent bounds on this shift \cite{pdg}. To satisfy all the limits, we take the mixing angle to be $s_\varphi = 10^{-2}$. Additionally, let $m_{H_1}=125$ GeV, $m_t=173$ GeV, $v=246$ GeV, and $v/\La=0.1$. As appropriate for $s_\varphi\simeq (\la_3/2\la_2)(v/\La)$, we derive $\la_3\simeq 0.1(m^2_{H_2}/\La^2)\sim 0.1$, since $2\la_2\simeq m^2_{H_2}/\La^2$ is to be taken as of order unity. As mentioned, we can put $h_{a3}=0.1$, as appropriate for the neutrino mass constraint. We first plot the dark matter annihilation cross sections to quarks, leptons, gauge bosons, and Higgs bosons as functions of the dark matter mass in Fig. \ref{fig4} for $m_{H_2}=2$ TeV, for example. We have taken $m_{\chi^0}=m_{\chi^-}=5$ TeV so that they are larger than the dark matter mass, as required. It is stressed that the choice of the next-to-lightest dark masses and the couplings/parameters above does not alternate the physical picture presented in this work. It is clear from Fig. \ref{fig4} that when $M_3<m_{H_2}$, the annihilation $\nu_{3R}\nu_{3R}\to H_2 H_2$ is kinematically suppressed, whereas when $M_3>m_{H_2}$, the annihilation channel is opened and larger than the other annihilation channels to quarks, leptons, gauge bosons, and $H_1$'s. In spite of this fact, in the weak coupling regime, i.e. $M_3/\La=-f_3/\sqrt{2} \lesssim \mathcal{O}(1)$ or $M_3\lesssim \mathcal{O}(1)\La\sim$ a few TeV, as supplied in the figure, only the $H_2$ resonance around 1 TeV is sufficiently strong to set the correct dark matter relic density, which requires $\langle \sigma v_{\mathrm{rel}}\rangle_{\nu_{3R}} \sim 1$~pb. Concretely, we plot the relic density $\Om_{\nu_{3R}} h^2\simeq 0.1\ \mathrm{pb}/\langle \sigma v_{\mathrm{rel}}\rangle_{\nu_{3R}}$ as a function of the dark matter mass in Fig.~\ref{fig5}. It is clear from Fig. \ref{fig5} that the $H_2$ resonance $m_{\nu_{3R}}=\fr 1 2 m_{H_2}$ is crucial to set the dark matter relic density for which the correct relic density acquires $M_3=0.98$--1.02 TeV.   

\begin{figure}[h]
\bc
\includegraphics[scale=0.8]{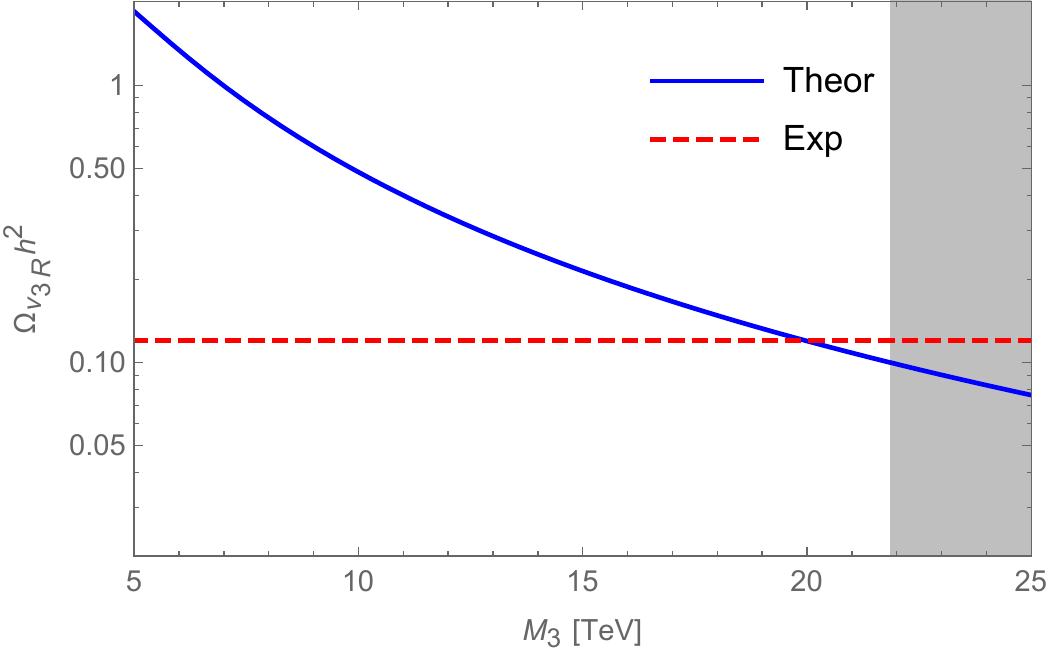}
\caption[]{\label{fig6} Majorana dark matter relic density set by its Yukawa coupling strength to the new Higgs boson before being suppressed by the unitarity condition, where the dashed line is the experimental value $\Om_{\mathrm{DM}} h^2 \simeq 0.12$ \cite{pdg}.}
\ec
\end{figure}

Exceptionally, as we can see in Fig. \ref{fig4}, the annihilation channel to $H_2H_2$ starts enhancing when $M_3$ is raised, i.e. the relevant coupling $f_3$ enters the strong regime. We plot the relic density in Fig. \ref{fig6} for large $M_3$ until it reaches the unitarity bound $M_3/\La=-f_3/\sqrt{2}=4\pi/\sqrt{2}$, or $M_3=21.85$ TeV, for $\La=10v=2.46$ TeV. The previous parameter values have been taken, except $m_{\chi^0}$ and $m_{\chi^-}$ would be chosen so that they are larger than the dark matter mass (their detailed values are insignificant). It is noted that the gray regime is excluded by the unitarity bound. That said, we can impose $M_3=19.9$--21.85 TeV so that the Majorana dark matter gets the correct density. Surely, the scotoseesaw still works for this high regime of dark matter mass. Additionally, in this case, $\nu_{3R}$ may annihilate to $Z'$ since $m_{Z'}=2g_D\La$ is radically smaller than the dark matter mass, where $g_D$ is the $U(1)_D$ gauge coupling. However, the relevant cross section is evaluated as $\langle \sigma v_{\mathrm{rel}}\rangle_{\nu_{3R}\nu_{3R} \to Z'Z'}\sim \al^2_D/M^2_3$, which negligibly contributes for $\al_D=g^2_D/4\pi<0.1$, similar to other channels to quarks, leptons, gauge bosons, and $H_1$'s. There are two available regimes for dark matter mass, but in the following we only consider the lower regime at TeV. Additionally, in this case, since the $t,u$-channel diagrams exchanged by dark fields infinitesimally contribute to the relic density set by the $s$-channel resonance, we need only require the dark fields to be heavier than the dark matter, i.e. $m_{R_{1,2}},m_{I_{1,2}},m_{C_{1,2}},M_{2}>M_{3}\sim 1$ TeV, which would be used for the subsequent phenomenology, similar to the neutrino sector.  

The effective Lagrangian describing scattering of $\nu_{3R}$ with quarks is induced directly from Fig. \ref{fig3} to be $\mathcal{L}_{\mathrm{eff}}\supset \al^S_q (\bar{\nu}^c_{3R}\nu_{3R})(\bar{q}q)$, where 
\be \al^S_q=-\fr{s_\varphi c_\varphi m_q M_3}{v \La }\left(\fr{1}{m^2_{H_1}}-\fr{1}{m^2_{H_2}}\right),\ee where the transfer momentum \cite{dmdd1} is much smaller than $m^2_{H_{1,2}}$ and is thus omitted. Further, because of $(1/m^2_{H_2})/(1/m^2_{H_1})=(m_{H_1}/m_{H_2})^2\sim 4\times 10^{-3}\ll 1$ for $m_{H_1}=125$ GeV and $m_{H_2}=2$ TeV as given above, the contribution of $H_2$ to the scattering process, i.e. the term $1/m^2_{H_2}$, is negligible and being suppressed, as aforementioned. Hence, the scattering cross section of dark matter with a nucleon $N=p,n$ is given by \cite{dmdd1}
\be \sigma^{\mathrm{SI}}_{\nu_{3R}-N}=\fr{4 m^2_N}{\pi} (f^N)^2,\ee where the nucleon form factor obeys 
\be \fr{f^N}{m_N}=\sum_{q=u,d,s}\fr{\al^S_q}{m_q} f^N_{Tq} +\fr{2}{27} f^N_{TG}\sum_{q=c,b,t} \fr{\al^S_q}{m_q}\simeq -\fr{0.35s_\varphi c_\varphi M_3}{v\La m^2_{H_1}},\ee where we have used $f^N_{TG}=1-\sum_{q=u,d,s}f^N_{Tq}$, $f^p_{Tu}=0.02$, $f^p_{Td}=0.026$, $f^p_{Ts}=0.118$, $f^n_{Tu}=f^p_{Td}$, $f^n_{Td}=f^p_{Tu}$, and $f^n_{Ts}=f^p_{Ts}$ \cite{dmdd2}. It follows that 
\be \sigma^{\mathrm{SI}}_{\nu_{3R}-N}\simeq \fr{0.49 m^4_N s^2_\varphi M^2_3}{\pi v^2 \La^2 m^4_{H_1}}\simeq \left(\fr{s_\varphi}{10^{-2}}\right)^2\left(\fr{M_3}{\mathrm{TeV}}\right)^2\times 0.68\times 10^{-46}\ \mathrm{cm}^2,\ee where we have taken $m_N=1$ GeV, $m_{H_1}=125$ GeV, $v=246$ GeV, and $v/\La = 0.1$. The SI cross section predicted is in agreement with the latest data $\sigma^{\mathrm{SI}}_{\nu_{3R}-N}\sim 10^{-46}\ \mathrm{cm}^2$ \cite{lzexp} for $\nu_{3R}$ dark matter mass $M_3\sim$ TeV and usual-new Higgs mixing $s_\varphi\sim 10^{-2}$.      

\subsection{Scalar dark matter}

Since the dark scalars $R_{1,2}$ and $I_{1,2}$ are heavy at TeV, their dark matter phenomenology is quite similar to the Majorana dark matter interpreted above. Indeed, the lightest of dark fields, $R_1$ for instance, is now stabilized by $P_D$. In the early universe, it may annihilate to a pair of $Z'$, as well as of weak bosons, through $t,u$-channel diagrams or to usual fermions via $H_{1,2}$ portals. However, only the $H_2$ portal is enhanced due to the mass resonance that sets the $R_1$ relic density. Concerning direct detection, $R_1$ scatters with the nucleon via the standard model Higgs portal $H_1$. The new Higgs portal gives a negligible contribution, while the $Z$ portal is inaccessible due to the dark scalar mass splitting. 

\section{\label{bounds} Constraints}

This section investigates various constraints coming from both the theoretical and experimental sides, which potentially restrict the viable parameter space of the model. 

\subsection{Theoretical constraint}

Besides the condition for vacuum stability briefly discussed in the scalar potential section, we force the scalar potential to be perturbative by requiring that all quartic couplings of the scalar fields satisfy $\la \leq 4\pi$, namely
\bea &&\left\{6|\la_1|,\ 6|\la_2|,\ |\la_3|,\ 6|\la_4|,\ 6|\la_5|,\ |\la_6|,\ |\la_6+\la_7|,\ |\la_8|,\right.\crn
&&\left.|\la_8+\la_{10}|,\ |\la_9|,\ |\la_9+\la_{11}|,\ |\la_{12}|,\ |\la_{13}|,\ |\la_{14}|\right\} \leq 4\pi.\eea

Additionally, we demand that the perturbative unitarity must be preserved in all scattering processes involving scalar and/or gauge fields. According to the Goldstone boson equivalence theorem, the gauge bosons behave as their Goldstone bosons at high energy. Hence, the scattering amplitudes of scalar and/or gauge fields at high energy match those that replace the gauge fields by their Goldstone bosons. In other words, it is sufficient to consider only the pure scalar scattering amplitudes that take scalar fields of both kinds, Goldstone and Higgs, into account. Because the trilinear scalar couplings negligibly contribute to high-energy scattering processes as suppressed by dimensional count, or transferred high energy, the perturbative unitarity demands that the scattering amplitudes take only quartic couplings of scalar fields into account, obeying $|\mathcal{M}|<8\pi$. For this aim, we can consider the scattering amplitudes of original scalar fields in the gauge bases because they can be related to the amplitudes in mass eigenstate bases by a unitary transformation and acquire that the eigenvalues of the scattering amplitude matrix obey $|\La_i|<8\pi$ \cite{peruni}. 

The $S$-matrix must conserve the exact symmetries presented in the model, such as electric charge $(Q)$, dark parity ($P_D$), and CP. Hence, the scattering amplitude matrix can be given in bases as 1) $Q=\pm 1$, $P_D$-even; 2) $Q=\pm 1$, $P_D$-odd; 3) $Q=0$, $P_D$-even, CP-even; 4) $Q=0$, $P_D$-odd, CP-even; 5) $Q=0$, $P_D$-even, CP-odd; 6) $Q=0$, $P_D$-odd, CP-odd. Taking, for instance, the explicit basis in case 2) is $(H^+R$, $H^+I$, $H^+R'$, $H^+I'$, $\eta^+S$, $\eta^+A$, $\eta^+S'$, $\eta^+A'$, $\chi^+S$, $\chi^+A$, $\chi^+ S'$, $\chi^+A')$. The scattering amplitude matrix is then
\bea \mathcal{M}_{Q=\pm1,P_D=-1}=\left(\begin{array}{cccccccccccc} 
\la_8 & 0 & 0 & 0 & \fr 1 2 \la_{10} & \fr 1 2 \la_{10} & 0 & 0 & -\fr 1 2 \la_{14} & -\fr{i}{2}\la_{14} & 0 & 0 \\
& \la_8 & 0 & 0 & \fr 1 2 \la_{10} & \fr 1 2 \la_{10}  & 0 & 0 & -\fr{i}{2}\la_{14} & \fr 1 2 \la_{14} & 0 & 0 \\
& & \la_9 & 0 & -\fr 1 2 \la_{14} & -\fr{i}{2}\la_{14} & 0 & 0 & \fr 1 2 \la_{11} & \fr 1 2 \la_{11} & 0& 0\\
& & & \la_9 & -\fr{i}{2}\la_{14} & \fr 1 2 \la_{14} & 0 & 0 & \fr 1 2 \la_{11} & \fr 1 2 \la_{11} & 0&0\\
& & & & \la_8 & 0 & 0 & 0 & 0 & 0 & 0& 0\\
& & & & & \la_8 & 0 & 0 & 0& 0& 0&0\\
& & & & & & \la_{12} & 0 & 0& 0& 0 & 0\\
& & & & & & & \la_{12} & 0& 0&0 & 0 \\
& & & (H.c.) & & & & & \la_9 & 0 & 0& 0 \\
& & & & & & & & & \la_9 & 0 & 0\\
& & & & & & & & & & \la_{13} & 0\\
& & & & & & & & & & & \la_{13} 
  \end{array}\right).\eea It is straightforward to write down the remaining amplitudes. Let us attract the reader's attention to Ref. \cite{aapara} for more details on an alternative model.              

\subsection{Precision electroweak measurement}

The new physics is associated with $U(1)_D$ gauge symmetry, right-handed neutrinos $\nu_{1,2,3R}$, and new Higgs fields $\phi,\eta,\chi$. When $\phi$ develops a VEV, $\langle \phi\rangle=\La/\sqrt{2}$, the $U(1)_D$ gauge boson $Z'$ obtains a mass $m_{Z'}=2g_D\La$ at $\La$ scale, where $g_D$ is the $U(1)_D$ coupling constant. 

It is clear that the gauge symmetry breaking does not induce a $Z$-$Z'$ mixing, since $\phi$ has only dark charge, whereas $H$ has only normal charges. Further, we assume that the kinetic mixing between $U(1)_Y$ and $U(1)_D$ gauge bosons, i.e. $B$ and $Z'$, respectively, vanishes at tree level, since this kind of mixing is possibly suppressed by a more fundamental gauge unification, such as GUT, which includes both $U(1)$'s in a simple group. 

Hence, the kinetic mixing is only induced by loop effects mediated by new fields that have both normal and dark charges~\cite{kimix}. In the present model, the inert scalar doublets $\eta,\chi$ play the role, inducing a kinetic mixing term, i.e. $\mathcal{L}\supset -(\epsilon/2)B^{\mu\nu}Z'_{\mu\nu}$, estimated naively as~\cite{kimixscalar} \be \epsilon\sim \fr{g_Y g_D}{48\pi^2}\ln \fr{m^2_\eta}{m^2_\chi}\sim 10^{-4}\left(\fr{g_D}{0.1}\right)\ln \fr{m^2_\eta}{m^2_\chi}\sim 10^{-4},\ee given that $g_D\sim 0.1$, where the $\eta$ ($\chi$) doublet, as proven before, is near degenerate, with a common mass $m_{\eta}$ ($m_{\chi}$) at $\La$ scale. This translates to a $Z$-$Z'$ mixing angle proportional to~$\epsilon$. 

The precision $Z$-pole physics is mainly sensitive to $Z$-$Z'$ mixing, which lowers the $Z$ mass and shifts $Z$-fermion couplings. Notice that $Z$-fermion couplings would determine various $Z$-pole observables, e.g. asymmetries and partial widths. The precision data bound the $Z$-$Z'$ mixing angle to be $\epsilon \lesssim 10^{-3}$, which is obviously satisfied (even for $g_D\sim 1$ instead) \cite{pdg}. The $Z$-mass shift leads to a deviation in the $\rho$-parameter at tree level, such as $\Delta\rho=m^2_W/c^2_W m^2_Z -1\sim \epsilon$, in agreement with the global fit $\Delta\rho=0.00031\pm 0.00019\sim 10^{-4}$ \cite{pdg}.

Besides, the low-energy weak neutral current experiments are potentially modified by the $U(1)_D$ gauge boson ($Z'$) exchange, mainly sensitive to its mass and couplings to fermions. However, since the usual fermions have no dark charge, the $Z'$-fermion couplings are suppressed by $\epsilon$, compared to those of $Z$. That said, $Z'$ does not modify the low-energy neutral currents for its mass beyond a sub-weak scale, i.e. $m_{Z'}\gg \epsilon m_Z\sim 10$ MeV, as desirable. 

Similar to the phenomenology of the usual and new gauge boson mixing, the $U(1)_D$ symmetry breaking field $H_2$ mixes with the standard model Higgs field $H_1$ via a mixing angle $\varphi\sim \fr{\la_3}{\la_2}\fr{v}{\La}$, as given above. This mixing potentially modifies the standard model Higgs couplings, which are well established at present. Comparing the effective Higgs couplings with standard model particles to existing data implies $\varphi \lesssim 10^{-2}$ \cite{pdg}, which has already been taken in the dark matter phenomenology.             

Last, but not least, the presence of inert Higgs doublets $\eta,\chi$ would contribute to self-energies of weak gauge bosons, given via oblique parameters $S,T,U$ \cite{peskinstu}. Because (i) the dark scalars $\eta,\chi$ do not mix with the usual Higgs field $H_1$ and the new Higgs field $H_2$ and (ii) the dark scalar mixings $\theta_C, \theta_R, \theta_I$ are very small, suppressed by $\la_{14}(v/\La)^2\ll 1$, the doublets $\eta,\chi$ and their component fields separately contribute to $S,T,U$. Actually, $U$ is very suppressed, compared to $S,T$. Hence, we study only $S,T$, given by \cite{stuidm} 
\bea S &=& \fr{1}{4\pi} \sum_{i=1,2}\left[\mathcal{G}(m^2_{R_i},m^2_{I_i})-\mathcal{G}(m^2_{C_i},m^2_{C_i})\right],\\ T &=& \fr{1}{16\pi^2 \al v^2}\sum_{i=1,2}\left[\mathcal{F}(m^2_{C_i},m^2_{R_i})+ \mathcal{F}(m^2_{C_i},m^2_{I_i})-\mathcal{F}(m^2_{R_i},m^2_{I_i})\right],\eea where \bea \mathcal{F}(x^2,y^2) &=& \fr{x^2+y^2}{2}-\fr{x^2y^2}{x^2-y^2}\ln \fr{x^2}{y^2}\crn
&\simeq& \fr 2 3 (x-y)^2,\\
\mathcal{G}(x^2,y^2)&=&-\fr 1 3 \left[\fr 4 3 -\fr{x^2\ln x^2 -y^2\ln y^2}{x^2-y^2}-\fr{x^2+y^2}{(x^2-y^2)^2}\mathcal{F}(x^2,y^2)\right]\crn
&\simeq & \fr 1 3 \ln y^2+\fr{x-y}{6y}.\eea Here, the approximations are given due to degenerate masses, i.e. $x\simeq y$. We derive
\bea T &\simeq& \fr{1}{12\pi^2\al v^2}\sum_{i=1,2}(m_{C_i}-m_{R_i})(m_{C_i}-m_{I_i}),\\
S&\simeq& -\fr{1}{6\pi}\sum_{i=1,2}\left(\fr{m_{C_i}-m_{I_i}}{m_{C_i}}-\fr{m_{R_i}-m_{I_i}}{4m_{I_i}}\right).\eea   
Because of $m_{R_i}-m_{I_i}\sim v^4/\La^3$ and $m_{C_i}-m_{R_i,I_i}\sim v^2/\La$, we further estimate 
\be S\sim -0.053(v/\La)^2,\hs  T\sim 1.08(v/\La)^2,\ee which predicts $S\sim -0.0005$ and $T\sim 0.01$ for $v/\La=0.1$, as taken above. This prediction agrees with the precision data, which constrain the new physics contribution to $S,T$ as
\be S=-0.05\pm 0.07,\hs T=0.00\pm 0.06,\ee for fixing $U=0$ \cite{pdg}.

Besides contributing to the oblique parameters, $\eta$ and $\chi$ would also mediate the standard model Higgs decay to two photons via one-loop effects, which potentially modifies their signal strength at the LHC, i.e. $\sigma(pp\to H_1\to \ga\ga)$. The signal strength of this model compared to that in the standard model is given as \be \mu_{\ga\ga} = \fr{\sigma(pp\to H_1)\mathrm{Br}(H_1\to \ga\ga)}{\sigma(pp\to H_1)_{\mathrm{SM}}\mathrm{Br}(H_1\to \ga\ga)_{\mathrm{SM}}}=\fr{\mathrm{Br}(H_1\to \ga\ga)}{\mathrm{Br}(H_1\to \ga\ga)_{\mathrm{SM}}},\ee where we use the narrow width approximation, and note that the Higgs production cross sections in this model and in the standard model are the same. [Exactly, they differ a factor close to unity, i.e. $c^2_\varphi \simeq 1$, due to the Higgs ($H_1$-$H_2$) mixing but suppressed by $\varphi\lesssim 10^{-2}$, as given. Hereafter, this mixing effect could be omitted.] Because the new fields $\eta,\chi,\phi, Z'$ are heavy, beyond the weak scale, the Higgs $H_1$ decays only to the standard model particles or, in other words, the total $H_1$ widths in this model and in the standard model are almost the same. We further have 
\be \mu_{\ga\ga}=\fr{\Ga(H_1\to \ga\ga)}{\Ga(H_1\to \ga\ga)_{\mathrm{SM}}}= \left|1+\fr{v^2}{2}\fr{\fr{\la_8+\la_{10}}{m^2_{C_1}}A_{C_1}+\fr{\la_9+\la_{11}}{m^2_{C_2}}A_{C_2}}{\fr{4}{3}A_t + A_W}\right|^2,\ee with the aid of \cite{higgsdecay}. Here, the decay $H_1\to \ga\ga$ is dominantly contributed by top and $W$ loops in the standard model, while it additionally includes contributions of $C^\pm_1\simeq \eta^\pm$ and $C^\pm_2\simeq \chi^\pm$ in the present model, where the charged dark Higgs mixing $\theta_C$ is small and omitted. The amplitudes are given by
\bea A_t &=& 2[\tau_t + (\tau_t-1)\arcsin^2(\sqrt{\tau_t})]/\tau^2_t\simeq 1.376,\\ 
 A_W &=& -[2\tau^2_W+3\tau_W+3(2\tau_W-1)\arcsin^2(\sqrt{\tau_W})]/\tau^2_W\simeq -8.324,\\
A_S &=& -[\tau_S - \arcsin^2(\sqrt{\tau_S})]/\tau^2_S\simeq \fr 1 3 +\fr{\tau_S}{36},\eea where $\tau_t=m^2_{H_1}/(4m^2_t)\simeq 0.13$, $\tau_W=m^2_{H_1}/(4m^2_W)\simeq 0.6$, and $\tau_S=m^2_{H_1}/(4m^2_S)\ll 1$, for $S=C_1,C_2$. Notice that $(\la_8+\la_{10})v$ and $(\la_9+\la_{11})v$ are the couplings of $H_1 C^+_1 C^-_1$ and $H_1 C^+_2 C^-_2$, respectively. We define a typical charged dark coupling and mass so that \be \fr{\bar{\la}}{m^2_C}\equiv \fr{\la_8+\la_{10}}{m^2_{C_1}}+\fr{\la_9+\la_{11}}{m^2_{C_2}}.\ee Since $\tau_S\ll 1$ and $(v/m_{C_{1,2}})^2\ll  1$, we approximate
\be \mu_{\ga\ga}\simeq 1-0.05\bar{\la}\left(\fr{v}{m_C}\right)^2.\ee Since $|\bar{\la}|<4\pi$ due to perturbative theory and $(v/m_C)^2\lesssim 0.06$ for $m_C\gtrsim 1$ TeV as already taken, we derive $0.961<\mu_{\ga\ga}<1.038$, in good agreement with the latest data $\mu_{\ga\ga}=1.02^{+0.09}_{-0.12}$~\cite{higgslhc}. In other words, the diphoton production strength (by Higgs decay) at the LHC is fulfilled for the viable parameter regime.            

\subsection{Lepton flavor violation}

The second and third terms in the Yukawa Lagrangian (\ref{eqadd121}) give rise to lepton flavor violation processes at the one-loop level, such as $e_a\to e_b\gamma$, $e_a\to 3 e_b$, and $\mu$-$e$ conversion in nuclei, where we assume the charged leptons $e_a/e_b$ by themselves to be flavor diagonal, i.e. physical fields $e,\mu,\tau$ according to $a/b=1,2,3$, respectively. 

Generally, a study of the current bounds for such processes would suggest that given the constraint on $e_a\to e_b\gamma$, the remaining processes are automatically satisfied (see, e.g., \cite{ahriche}). Hence, we consider only the process $e_a \to e_b\gamma$, governed by 
\bea \mathcal{L} &\supset& h_{a2} \bar{e}_{aL}\nu_{2R}(c_C C^-_1+s_C C^-_2)+h_{a3}\bar{e}_{aL}\nu_{3R}(c_C C^-_2-s_C C^-_1)\crn 
&\simeq& h_{a2} \bar{e}_{aL}\nu_{2R} C^-_1+h_{a3}\bar{e}_{aL}\nu_{3R}C^-_2,\eea where we denote $c_C\equiv c_{\theta_C}$ and $s_C\equiv s_{\theta_C}$, similar to those in the neutral dark sector. Additionally, because of $\theta_C\sim \la_{14}(v/\La)^2\ll 1$, the relevant interactions negligibly contribute to the flavor changing, as omitted. In other words, we can identify $C^-_1\simeq \eta^-$ and $C^-_2\simeq \chi^-$.

Generalizing the result in \cite{vicente}, we get the branching ratio, 
\be \mathrm{Br}(e_a\to e_b \gamma)=\fr{3\al v^4}{32\pi}\left|\fr{h_{a2}h^*_{b2}}{m^2_{C_1}}F\left(\fr{M^2_2}{m^2_{C_1}}\right)+\fr{h_{a3}h^*_{b3}}{m^2_{C_2}}F\left(\fr{M^2_3}{m^2_{C_2}}\right)\right|^2\mathrm{Br}(e_a\to e_b\nu_a\bar{\nu}_b),\ee where the loop function is given by $F(x)=(1-6x+3x^2+2x^3-6x^2\ln x)/6(1-x)^4$. This loop function has no pole at $x=1$, decreasing when $x$ increases from zero, thus bounded by $F(x)<1/6$ for $x>0$. Therefore, the most severse process obeys 
\be \mathrm{Br}(\mu\to e\gamma)\lesssim 2.37\times 10^{-14}\left(\fr{|h_{\mu k}h^*_{e k}|}{10^{-3}}\right)^2\left(\fr{1\ \mathrm{TeV}}{m_{C_{1,2}}}\right)^4,\ee with the aid of $m_{C_1}\sim m_{C_2}$, $\al=1/128$, $v=246$ GeV, and $\mathrm{Br}(\mu \to e \nu_\mu \bar{\nu}_e)\simeq 1$, where $k=2,3$. This prediction is below the current bound $\mathrm{Br}(\mu\to e \gamma)\simeq 4.2\times 10^{-13}$ reported by MEG \cite{meg}, even agreeing with an upgrade for sensitivity at $\mathrm{Br}(\mu\to e \gamma)\sim 6\times 10^{-14}$ by this experiment \cite{meg2}, for $|h_{\mu k }h^*_{ek }|\sim 10^{-3}$ and $m_{C_{1,2}}\sim 1$ TeV, similar to those in the neutrino mass generation if taking $\la_{14}\sim h_{ak}$ and $m_{C_{1,2}}\sim M_{2,3}$.

\subsection{Collider signatures}

The new gauge and Higgs bosons $Z'$ and $H_2$ weakly couple to normal matter, as suppressed by mixing parameters $\epsilon \sim 10^{-4}$ and $\varphi \sim 10^{-2}$, respectively. Additionally, they possess a mass proportional to the new physics scale $\La$ at TeV. Hence, these new fields cannot be produced at the LEPII, as they are kinematically suppressed \cite{lepii}. Further, these new fields represent negligible signals, if created, compared with those bounded by the LHC experiments due to the weak couplings \cite{lhc1,lhc2,lhc3}. On the other hand, the right-handed neutrino $\nu_{1R}$ couples to charged leptons and neutrinos via the usual Higgs doublet to be rather weak, say $h_{a1}\sim \sqrt{m_\nu M_1}/v\sim 10^{-6}$, for the seesaw scale $M_1$, i.e. the $\nu_{1R}$ mass, at the TeV scale. This right-handed neutrino also interacts with normal particles through weak currents due to its mixing with usual neutrinos, suppressed by a mixing angle $\sim m_{a1}/M_1\sim 10^{-6}$ to be very small. That said, similar to $Z'$ and $H_2$, the signature of $\nu_{1R}$ at colliders---which would decay to charged leptons, usual Higgs bosons, or weak gauge bosons, where the last two subsequently decay to jets and leptons---is insignificant too due to the weak coupling~\cite{lhc4}.

We have signified the dark matter candidate, the lightest of the dark fields, to be heavy, with a mass in the TeV regime. Therefore, they cannot be generated at the LEPII and even Tevaron colliders, as kinematically forbidden. However, they may be produced at the LHC in the form of large missing transverse momentum (or energy, $E\!\!\!\!/_T$), since the dark scalar doublets directly couple to electroweak gauge (possibly Higgs) bosons. We divide this into two cases depending on which dark field is dark matter. 

If dark matter is a neutral dark scalar, say $R_1$, possible signatures at the LHC include \ben
\item $pp \to C^+_1 C^-_1\to W^+W^- R_1R_1$, where $C^+_1C^-_1$ are produced via $\ga,Z$ exchanges. The final states can be recognized in the form of 4 jets, $W^+W^-\to u d^c d' u'^c$, 2 jets plus 1 charged-lepton, $W^+W^-\to u d^c l \nu^c$, or dilepton $W^+W^-\to l^c\nu l'\nu'^c$, recoiled against large missing energy carried by dark matter and/or invisible neutrinos.
\item $pp\to C^\pm_1 I_1\to W^\pm Z R_1 R_1$, where $C^\pm_1 I_1$ are created by $W^\pm$ exchange. The final state signals 4 jets, $W^\pm Z\to (ud^c/du^c)qq^c$, 2 jets plus dilepton, $W^\pm Z\to (ud^c/du^c)ll^c$, 2 jets, $W^\pm Z\to (ud^c/du^c)\nu\nu^c$, 1 charged-lepton plus 2 jets, $W^\pm Z\to (l^c\nu/l\nu^c)qq^c$, or 1 charged-lepton, $W^\pm Z\to (l^c\nu/l\nu^c)\nu\nu^c$.  
\item $pp \to C^\pm_1 R_1\to W^\pm R_1 R_1$, where $C^\pm_1 R_1$ are created by $W^\pm$ exchange, for which the final state signals 2 jets, $W^\pm\to ud^c/du^c$, or 1 charged-lepton, $W^\pm\to l^c\nu/l\nu^c$.
\item $pp\to I_1R_1\to Z R_1 R_1$, where $I_1 R_1$ are produced by $Z$ exchange, which subsequently signals 2 jets, $Z\to qq^c$, or dilepton, $Z\to ll^c$.
\item $pp\to I_1I_1\to Z Z R_1 R_1$, where $I_1 I_1$ are directly generated by the Higgs exchange, which signals 4 jets, $ZZ\to qq^cq'q'^c$, 2 dileptons, $ZZ\to ll^cl'l'^c$, 2 jets plus dilepton, $ZZ\to qq^c ll^c$, or 2 jet, $ZZ\to qq^c\nu\nu^c$.
\item $pp\to X R_1 R_1$, where $R_1R_1$ is directly produced via the Higgs portal, which signals a mono-$X$ emitted from an incoming proton, where $X$ is a jet or a photon.
\item Numerous processes that initially produce the second dark scalar doublet ($C_2, I_2,R_2$) instead of the first dark scalar doublet ($C_1,I_1,R_1$) above, or a mixture of these two dark scalar doublets, before the intermediate dark fields decay to dark matter $R_1$'s plus jet or lepton signals.
\een   

If dark matter is a neutral dark fermion, say $\nu_{3R}$, possible signatures at the LHC contain
\ben
\item $pp\to C^+_2C^-_2\to ll^c \nu_{3R}\nu^c_{3R}$, realizing a dilepton signal $ll^c$, recoiled against large missing energy carried by dark matter $\nu_{3R}\nu^c_{3R}$.
\item $pp\to C^\pm_2 (I_2/R_2)\to (l^c\nu/l\nu^c)\nu_{3R}\nu^c_{3R}$, realizing a charged-lepton signal, besides large missing energy that includes both dark matter and invisible neutrino.  
\item $pp\to X(I_2 R_2/I_2I_2/R_2R_2)\to X\nu\nu^c\nu_{3R}\nu^c_{3R}$, realizing a mono-$X$ (jet or photon) signal, emitted from an initial proton.  
\item Various processes that produce either $(C_1,I_1,R_1)$ instead of $(C_2,I_2,R_2)$ above or a mixture of two kinds of dark scalars before the relevant dark scalars decay to dark matter plus lepton or mono-$X$ signals.   
\een

Particularly, the LHC experiments have investigated slepton-pair production then decaying to a dilepton signal $ll^c$ plus missing energy carried by neutralino dark matter, say $pp\to \tilde{l}\tilde{l}^*\to ll^c \tilde{\chi}^0_1 \tilde{\chi}^0_1$, making a bound for slepton mass to be $m_{\tilde{l}}>700$ GeV, assuming $\mathrm{Br}(\tilde{l}\to l \tilde{\chi}^0_1)=1$ \cite{susylhc1,susylhc2}. If one identifies the second dark scalar doublet to the slepton doublet, the $C_2$ production cross section matches that of slepton. Additionally, assuming $C_2$ lighter than every dark field except for dark matter $\nu_{3R}$ leads to $\mathrm{Br}(C_2\to l\nu_{3R})=1$. Thus, the SUSY result applies to our case, such as $m_{C_2}>700$ GeV, as expected.       

\section{\label{concl} Conclusion and outlook}

We have shown that if the three right-handed neutrinos---the right-handed partners of the usual left-handed neutrinos---possess a dark gauge charge while the standard model particles do not, their charges would be $0,-1,+1$ in order for the theory to be consistent at the quantum level. This new theory naturally implies the scotoseesaw mechanism which induces appropriate neutrino masses via both contributions of DM (scotogenic) and non-DM (seesaw) right-handed neutrinos. The dark matter candidate is either one of the dark right-handed neutrinos or one of the neutral dark scalars, obtaining a mass at TeV, possessing a correct abundance enhanced by the new Higgs mass resonance $m_{\mathrm{DM}}=\fr 1 2 m_{H_2}$, and potentially scattering with nucleons via the standard model Higgs exchange.          

Since the present work has not signified a kinetic mixing, or this mixing is small, $Z'$ does not significantly couple to usual fermions. Hence, the current colliders \cite{lepii,lhc1,lhc2} do not make a constraint on $Z'$. Similarly, the other new particles, such as DM ($\nu_{3R}$ or $R_1$) and non-DM ($\nu_{1R}$ and $H_2$), would exhibit negligible signals at the present colliders because they interact rather weakly with the normal matter. However, the projected HL-LHC and ILC may be worth exploring for them when the energy and sensitivity are enhanced.   

\section*{Acknowledgement}

This research is funded by Vietnam National Foundation for Science and Technology Development (NAFOSTED) under grant number 103.01-2023.50.

\appendix

\section{\label{appa} Verifying radiative corrections by the normal Higgs fields}

\begin{figure}[h]
\bc
\includegraphics[]{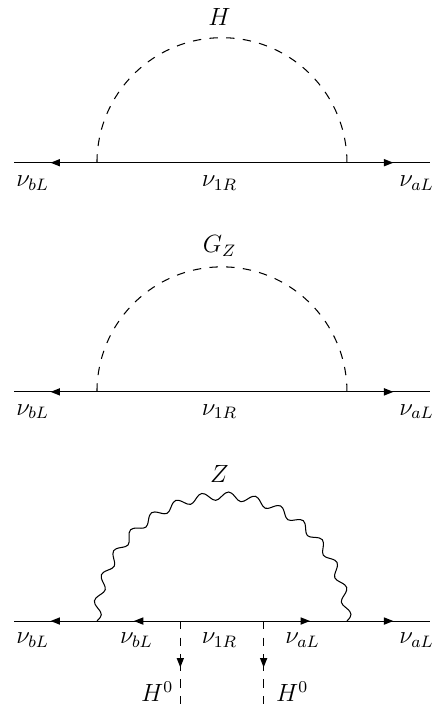}
\caption[]{\label{fig7} One-loop corrections of normal Higgs and gauge fields to neutrino masses.}
\ec
\end{figure}

The neutrino masses, as given in (\ref{adt1}), are induced by the scotoseesaw scheme, which includes both tree- and loop-level contributions in Fig. \ref{fig1}. However, at the one-loop level, there exist radiative corrections mediated by $H_{1,2}$, $G_{Z,Z'}$, and $Z,Z'$ that couple to neutrinos, which have not been accounted for. Since $H_2$, $G_{Z'}$, and $Z'$ couple only to neutrinos via small mixing effects, their contributions to neutrino masses are indeed negligible. The significant contributions come from $H_1$, $G_Z$, and $Z$, as depicted in Fig. \ref{fig7}, where we redefine $H\equiv H_1$, for brevity.\footnote{It is noted that in a generic $R_\xi$-gauge, the second diagram by $G_Z$ must contribute. Even in the unitarity gauge for which the second diagram disappears, the third diagram by $Z$ always contributes. An evaluation of radiative correction by the Higgs field to neutrino mass must include the relevant Goldstone boson and/or gauge boson, which has a nature different from those by scotogenic fields.} In other words, it is sufficient to verify the contributions of the standard model Higgs and $Z$ fields, without loss of generality. It is stressed that because (i) our theory is renormalizable and (ii) neutrino mass vanishes at the tree level, every radiative correction, including both dark fields, as in the body text, and normal fields, as in this appendix, to neutrino mass must be finite. In what follows, we consider the latter---that is mediated by the normal fields---and show that such contribution is radically smaller than the tree-level seesaw mass, which should be omitted. 

Applying the Feynman rules in the generic $R_\xi$-gauge, we derive the radiative neutrino mass in the form of $\mathcal{L}\supset -\fr 1 2 \bar{\nu}_{aL} (M_\nu)_{ab} \nu^c_{bL}+H.c.$, in which  
\bea -i(M_\nu)_{ab}P_R &=& \int \fr{d^4 p}{(2\pi)^4}\left(\fr{i}{\sqrt{2}}h_{a1}P_R\right)\fr{i}{p\!\!\!/  - M_1}\left(\fr{i}{\sqrt{2}}h_{b1}P_R\right)\fr{i}{p^2-m^2_H}\crn
&&+ \int \fr{d^4 p}{(2\pi)^4}\left(\fr{1}{\sqrt{2}}h_{a1}P_R\right)\fr{i}{p\!\!\!/  - M_1}\left(\fr{1}{\sqrt{2}}h_{b1}P_R\right)\fr{i}{p^2-\xi m^2_Z}\crn
&& +\int \fr{d^4p}{(2\pi)^4}\left(\fr{-ig}{2c_W}\ga^\mu P_L\right)\fr{i}{p\!\!\!/} \left(i h_{a1}\fr{v}{\sqrt{2}}P_R\right)\fr{i}{p\!\!\!/-M_1}\left(i h_{b1}\fr{v}{\sqrt{2}}P_R\right)\crn
&&\times \fr{i}{p\!\!\!/}\left(\fr{ig}{2c_W}\ga^\nu P_R\right)\fr{-i }{p^2-m^2_Z}\left(g_{\mu\nu}-\fr{(1-\xi)p_\mu p_\nu}{p^2-\xi m^2_Z}\right), \eea where the vertices are extracted from the Lagrangian $\mathcal{L}\supset h_{a1}\bar{\nu}_{aL}\nu_{1R}(v+H-i G_Z)/\sqrt{2}+H.c.-(g/2c_W)\bar{\nu}_{aL}\ga^\mu \nu_{aL}Z_\mu$, while we notice that the propagator of $G_Z$ takes the form $i/(p^2-\xi m^2_Z)$ in the generic $R_\xi$-gauge, since $G_Z$ is the Goldstone coupled to $Z$, as usual. 

It is straightforward to deduce 
\bea (M_\nu)_{ab} &=& \fr{ih_{a1}h_{b1} M_1}{2} \int\fr{d^4p}{(2\pi)^4}\left[\fr{m^2_H-m^2_Z}{(p^2-m^2_H)(p^2-m^2_Z)(p^2-M^2_1)}\right.\crn
&&\left.+\fr{4m^2_Z}{p^2(p^2-m^2_Z)(p^2-M^2_1)}\right],\eea which is independent of the gauge-fixing parameter $\xi$, as expected. This mass $M_\nu$ contributes directly to the tree-level seesaw mass in (\ref{dtttn1}) since $m_{a1} = -h_{a1}v/\sqrt{2}$, but not to the scotogenic mass as in (\ref{raddtttn}). $M_\nu$ is smaller than $m^{\mathrm{tree}}_\nu$ by a loop factor of $1/32\pi^2\sim 3\times 10^{-3}$, which is safely suppressed. Indeed, because of \bea I(a,b,c) &=& \int \fr{d^4 p}{(2\pi)^4}\fr{1}{(p^2-a)(p^2-b)(p^2-c)}\crn
&=&-\fr{i}{16 \pi^2}\left[\fr{a\ln a}{(a-b)(a-c)}+\fr{b\ln b}{(b-a)(b-c)}+\fr{c\ln c }{(c-a)(c-b)} \right],\eea we have \be I(0,b,c)\simeq -\fr{i}{16\pi^2}\fr{\ln c/b}{c},\ee for $c\gg b$, and 
\be I(a,b,c)\simeq -\fr{i}{16\pi^2}\fr{a\ln c/a -b\ln c/b}{(a-b)c},\ee for $c\gg a\sim b$.         
Applying these approximations for $M_1\gg m_{H,Z}$ to $M_\nu$, we get
\be (M_\nu)_{ab} \simeq \fr{h_{a1}h_{b1}}{M_1}\fr{m^2_H\ln M^2_1/m^2_H + 3 m^2_Z\ln M^2_1/m^2_Z}{32\pi^2}.\label{radss}\ee Because of $m_{H,Z}\sim v$ and $\ln M_1/m_{H,Z}\sim 1$, the radiative mass $M_\nu$ is suppressed by $1/32\pi^2$, compared to the tree-level seesaw mass $m^\mathrm{tree}_\nu$, as is desirable. 

The result in (\ref{radss}) agrees with that in \cite{radseesaw,radseesaw1}, which proved that radiative correction to seesaw neutrino mass is finite, as in our study.

\end{document}